\pdfoutput=0
\documentclass[10pt,twocolumn,conference]{IEEEtran}
\usepackage{amsmath,amssymb,amsthm}
\usepackage[noadjust]{cite}
\usepackage{gensymb}

\usepackage[font=footnotesize]{caption}
\usepackage{subcaption}
\usepackage{xspace,arydshln}
\usepackage{graphicx,fancyhdr,epstopdf}
\graphicspath{{./figures/}}
\usepackage{enumitem}
\usepackage{color}
\usepackage{hyperref}
\usepackage{soul}
\usepackage{etoolbox}

\input arraymac.tex
\usepackage{comment}
\usepackage{steinmetz} 
\usepackage{gensymb} 
\usepackage{mathrsfs}
\usepackage[normalem]{ulem} 
\usepackage{algorithm}
\usepackage{algpseudocode}
\DeclareMathOperator{\snr}{SNR}
\DeclareMathOperator{\slnr}{SLNR}
\newcommand{\round}[1]{\ensuremath{\lfloor#1\rceil}}

\newcommand\blfootnote[1]{%
  \begingroup
  \renewcommand\thefootnote{}\footnote{#1}%
  \addtocounter{footnote}{-1}%
  \endgroup
}

\begin{document}
\title{Design and Operation Principles of a Wave-Controlled Reconfigurable Intelligent Surface}
\author{\IEEEauthorblockN{Gal Ben Itzhak, {\em Graduate Student Member,~IEEE,}}
\IEEEauthorblockN{Miguel Saavedra-Melo, {\em Graduate Student Member,~IEEE,}}
\IEEEauthorblockN{Benjamin Bradshaw, {\em Graduate Student Member,~IEEE,}}
\IEEEauthorblockN{Ender Ayanoglu, {\em Fellow,~IEEE,}}
\IEEEauthorblockN{Filippo~Capolino,~{\em Fellow,~IEEE,}}
\IEEEauthorblockN{A. Lee Swindlehurst, {\em Fellow,~IEEE}}
}
\maketitle

\begin{abstract}
A Reflective Intelligent Surface (RIS) consists of many small reflective elements whose reflection properties can be adjusted to change\blfootnote{The authors are with the Center for Pervasive Communications and Computing (CPCC), Department of Electrical Engineering and Computer Science (EECS), University of California, Irvine, Irvine, CA 92697.}\blfootnote{This work si partially supported by the National Science Foundation grant 2030029.} the wireless propagation environment. Envisioned implementations require that each RIS element be connected to a controller, and as the number of RIS elements on a surface may be on the order of hundreds or more, the number of required electrical connectors creates a difficult wiring problem, especially at high frequencies where the physical space between the elements is limited. A potential solution to this problem was previously proposed by the authors in which ``biasing transmission lines" carrying standing waves are sampled at each RIS location to produce the desired bias voltage for each RIS element. This solution has the potential to substantially reduce the complexity of the RIS control. This paper presents models for the RIS elements that account for mutual coupling and realistic varactor characteristics, as well as circuit models for sampling the transmission line to generate the RIS control signals. For the latter case, the paper investigates two techniques for conversion of the transmission line standing wave voltage to the varactor bias voltage, namely an envelope detector and a sample-and-hold circuit. The paper also develops a modal decomposition approach for generating standing waves that are able to generate beams and nulls in the resulting RIS radiation pattern that maximize either the Signal-to-Noise Ratio (SNR) or the Signal-to-Leakage-plus-Noise Ratio (SLNR).
Extensive simulation results are provided for the two techniques, together with a discussion of computational complexity.
\end{abstract}
\begin{IEEEkeywords}
Spatial Fourier series, envelope detection, sample-and-hold, least squares (LS), simulated annealing (SA).
\end{IEEEkeywords} 
\section{Introduction}
Reconfigurable Intelligent Surface (RIS) technology provides controllable degrees-of-freedom (DoFs) for shaping the wireless radio-frequency (RF) channel in advantageous ways, for example by steering signals around blockages, providing beamforming gain to enhance signal-to-noise ratio (SNR) and reduce interference, and improving the overall quality-of-service (QoS) enjoyed by network users \cite{PanZhang22}. An RIS is populated by a typically large number of essentially passive (i.e., gainless) elements such as metallic patches whose reflective properties can be externally controlled. For an RIS with $R$ rows and $M$ elements per row, the total number of elements is defined as $M'=M\times R$. In common implementations, the reconfigurability is achieved by varying the biasing voltage across a varactor or the current through a p-i-n diode present in each element, which in turn produces variations in the input impedance seen by impinging RF energy. When properly designed, the electrical control can tune the reflection phase of each element in a particular frequency band to nearly any value between $-\pi$ and $\pi$. Some designs also provide tunability of the reflection amplitude to values between~0 and~1 (due to the element's passivity), although in many cases it is common to maintain the amplitude as close to unity as possible.

At millimeter wave or terahertz frequencies, an RIS can be designed with hundreds or potentially thousands of elements in a relatively small form factor, enabling large beamforming gains and narrow reconfigurable pencil-like beams. While having such high gains and directivity is advantageous, it comes with certain implementational challenges. First, because most RIS designs do not include active receivers, they must be controlled by an external device such as an access point or basestation (BS). This means that the wireless channels to/from an $M'$-element RIS must be estimated remotely at the BS, which can lead to an $M'$-fold increase in the pilot overhead unless certain assumptions are made about the propagation environment, such as the presence of only sparse propagation paths (reasonable at high frequencies) \cite{SwindleZLPL22}. Second, once the channel is estimated, the BS must estimate the optimal RIS configuration, which typically requires a complicated non-convex optimization over more than $M'$ variables, and then it must transmit the optimal configuration composed of $M'$ complex values to the RIS to control its behavior. Clearly, a large value for $M'$ will in turn create a large signaling overhead. This overhead is often manageable since the RIS need only be updated at the channel coherence rate, but techniques have nonetheless been proposed to compress the required amount of information flow using for example entropy coding \cite{XiaCY21} or by approximating the RIS phase vector using a low-rank tensor \cite{Sokal23}. A third more difficult challenge arising from large RIS with many elements is the apparent need for $M'$ wired connections to supply the required voltages or currents to all RIS elements. This requires an intricate design with potentially thousands of individual signal pathways throughout the device. Addressing this design issue has received considerably less attention, with some proposals suggesting the use of light-based controls \cite{Sayanskiy23,MiaoL23}.

To overcome these limitations, in this paper we propose an alternative technique that uses a {\em single\/} electric connection for each row of $M$ unit cells, as shown in Fig.~\ref{fig:RIS_Figure_1}, resulting in a reduced-dimension method for controlling the RIS element behavior that leads to both a simpler hardware implementation and a lower signaling overhead. Furthermore, we also provide an electromagnetic model to estimate analytically the reflection coefficient that accounts for mutual couplings and losses in the materials and includes a simple SPICE-based model of a commercially available varactor.

\section{Assumptions and Notation}
\begin{figure}[!t]
\begin{center}
  \includegraphics[width=3.0in]{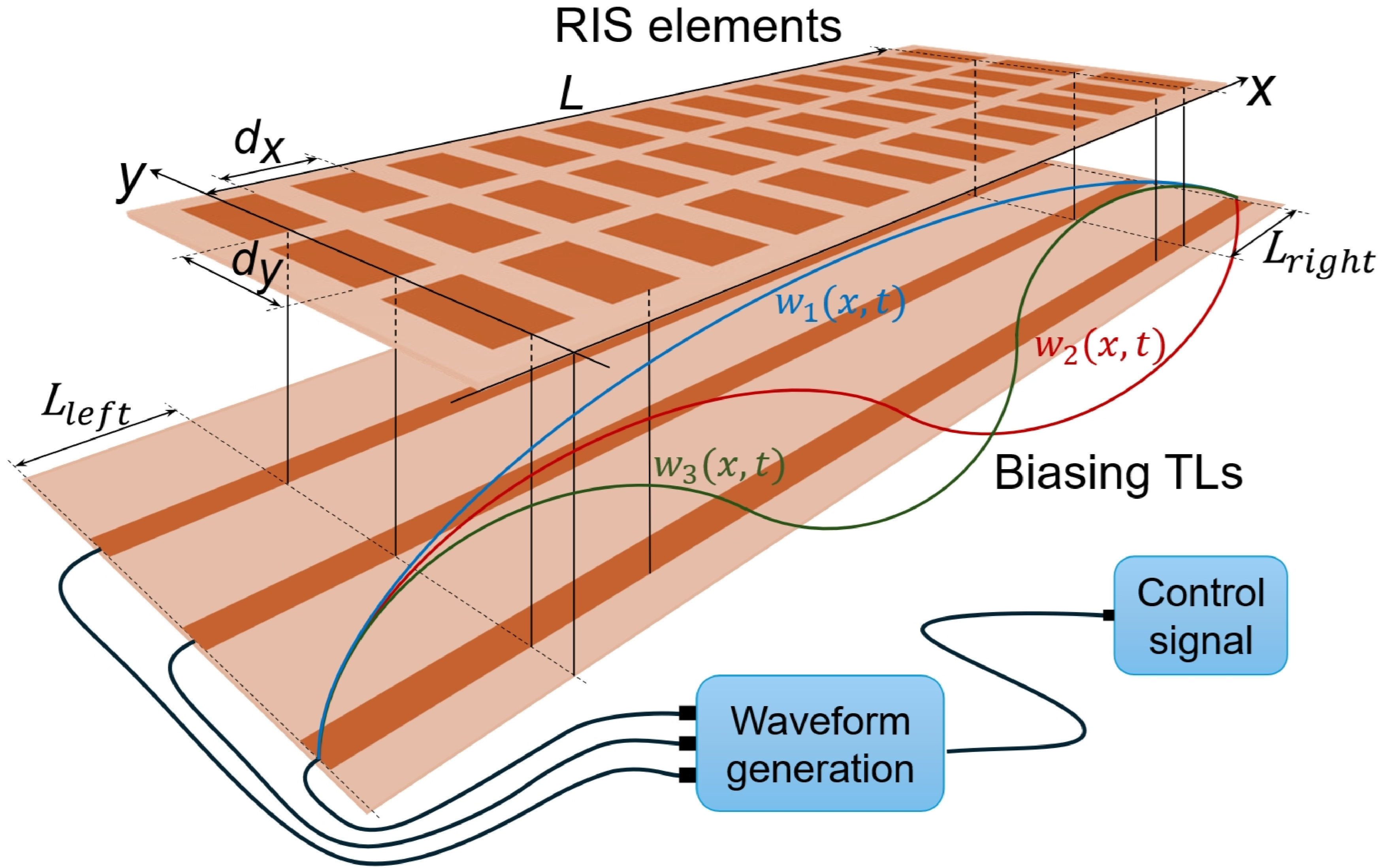}
  \caption{Wave-controlled RIS made of two physical layers. Top layer: $M$ RIS elements in each row along $x$; each element is connected to a varactor diode. Bottom layer: $N$ standing waves along the biasing transmission lines (TLs) to create the biasing voltages when sampled at each RIS element. Each row is controlled only by the connection at the left where $N$ frequencies are injected by a waveform generator.}\label{fig:RIS_Figure_1}
\end{center}
\end{figure}
We assume that each row (or column) of the RIS has $M$ metallic patches connected with vertical vias to a waveguide (located at a lower level) excited in such a way to introduce $N$ standing waves from which the biasing voltage for each element in the row can be induced \cite{main-ris}. As shown in Fig.~\ref{fig:RIS_Figure_1}, the
standing waves are assumed to be parameterized by $N \ll M$ harmonic modes, whose coefficients form the control information that defines the RIS response. A large reduction in degrees of freedom is foreseen for RIS that have a large number $M$ of elements along $x$, for example on the order of hundreds or even thousands. Thus, to configure the RIS response, the BS can perform an optimization over a much smaller number $N$ of parameters, and transmit a much smaller set of data to the RIS for its control. Moreover, the need for dense wiring and signal paths that would be required to physically connect to every RIS element is avoided, while still guaranteeing a large degree of control of the RIS. This offers a substantial reduction in the required hardware that is particularly important at millimeter waves where the physical space is limited.

We note here that the proposed architecture is different from that for so-called Dynamic Metasurface Antennas (DMAs) \cite{SIXHDRS16,NirDMA21}, which also employ waveguides along the rows or columns of the metasurface to connect to the individual elements. However, in a DMA, the waves entering the surface at each element combine together and propagate along the waveguide before being sampled for processing. This allows for an active implementation with (for example) signal amplification, but the beamforming must take into account the inherent analog combining that occurs in the waveguide. The operation of our proposed design is more akin to a conventional RIS, the key difference being how the control signals for each RIS element are generated.

To describe the performance of our proposed wave-controlled approach, we first provide a detailed model for an RIS design based on varactor diode control, and verify the accuracy of the model using full-wave electromagnetic simulations. The model accounts for mutual coupling among the RIS elements. It also incorporates realistic non-ideal behavior due to losses in the metallic patches, in the dielectric substrate, and in the varactor diodes, leading to realistic voltage- and frequency-dependent variations in the RIS element reflection coefficient amplitudes and phases \cite{Hanna22}. We also discuss methods to interface the waveguide control with the proposed RIS unit cells.
We will present several numerical examples involving a reflective metasurface implementation to compare three different ways to control the RIS, namely: (i) \textit{Ideal Phase} -- The reflection phases of the elements are perfectly tuned; (ii) \textit{Arbitrary Voltage Bias} -- Each varactor is biased using an arbitrary voltage to create the reflection magnitude and phase based on the analytical model of the RIS elements; (iii) \textit{Wave-Controlled Bias} -- The standing waves are used to control the varactors and reflection coefficients.
The results demonstrate the ability of the reduced dimension parametric control implemented with a realistic RIS to achieve performance close to that obtained in the idealized cases.

We consider a narrowband flat fading scenario with a single-antenna transmitter (Tx), $K$ single-antenna receivers (Rx), and an RIS with $M$ elements. To focus on the behavior of the RIS, in this work we will assume there is no direct signal path between the Tx and Rx. In this case, the signal $y_k$ at the $k$-th Rx will be given by the following signal expression assuming a transmitted signal $s$:
\begin{equation}
    y_k=\boldsymbol{h}_k^T\boldsymbol{\Phi}\boldsymbol{g} s + n_k \; ,
\end{equation}
where $n_k$ represents noise, $\hbf_k$ and $\gbf$ are respectively the $M \times 1$ channels from the RIS to the $k$-th Rx and the Tx to the RIS. The RIS response is defined by a diagonal matrix whose elements contain the reflection coefficients at the RIS elements:
\begin{equation}
    \boldsymbol{\Phi}=\text{diag} \left[ \phi(0), \phi(1), \ldots , \phi(M-1) \right] .
\end{equation}
As described in the next section, in a varactor-based implementation, the value of the $m$-th reflection coefficient $\phi(m)$ is determined by a biasing voltage applied to the $m$-th RIS element. Due to the passive nature of each element, the reflection coefficients satisfy $|\phi(m)| \le 1$ for all $m=0,1,\ldots,M-1$. In the following, we will let $\phibf=[\phi(0), \phi(1), \ldots, \phi(M-1)]^T$ denote the vector comprising the RIS reflection coefficients.
The achievable values for the reflection coefficients as a function of frequency and varactor bias voltage are determined by considering mutual coupling under the local periodicity condition, as explained in \cite{Hanna22} and also studied in \cite{Costa21, Nayeri18Ch10}.

%
\section{Varactor-Based RIS Reflection Model}\label{sec:model}
\begin{figure}[!t]
\begin{center}
  \includegraphics[width=3.0in]{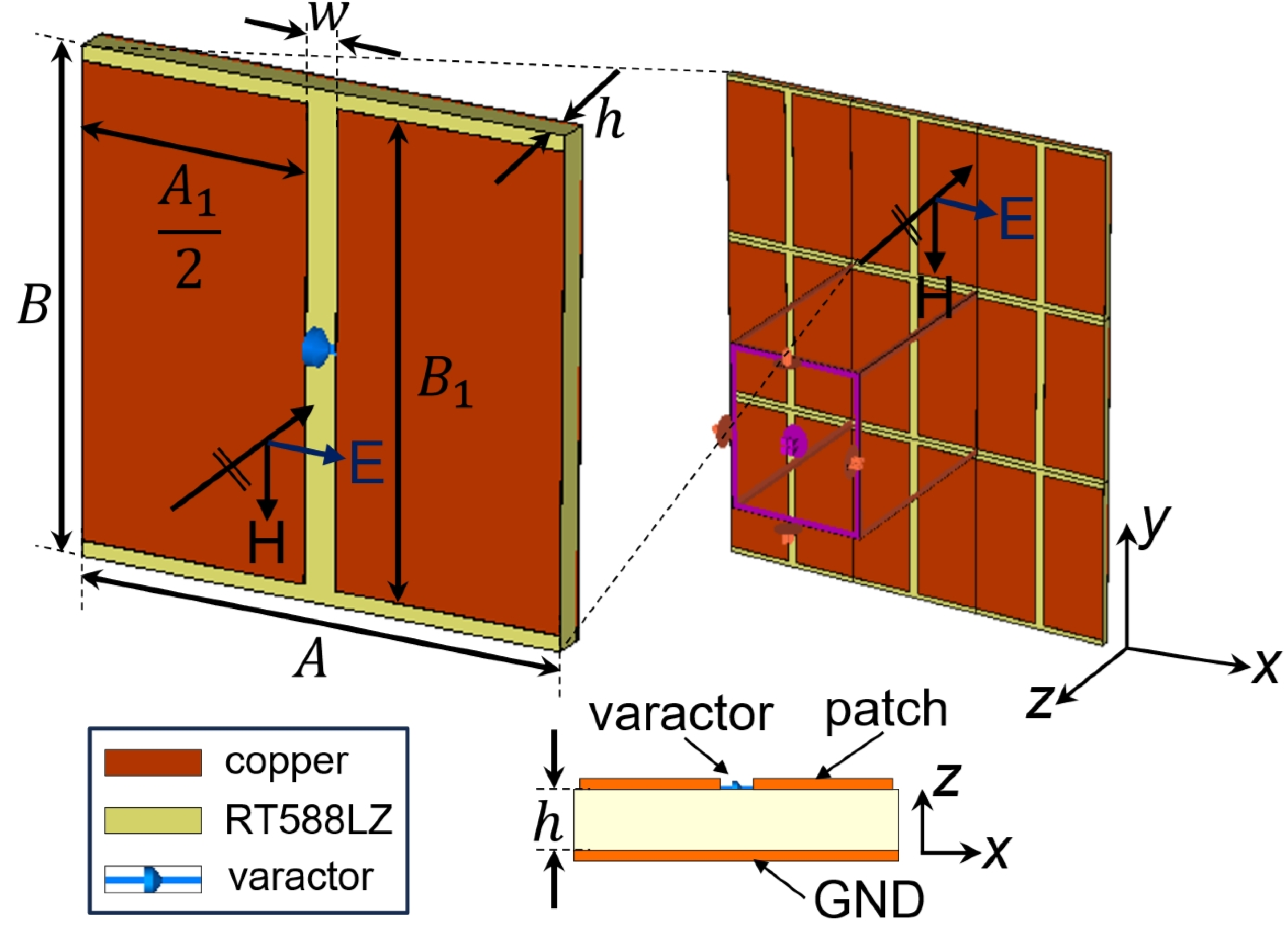}
  \caption{RIS formed by a periodic arrangement of square metallic conductors on a grounded dielectric substrate. The polarization of the incident electric field is along $x$. Varactor diodes are between patches, used as tunable capacitors when reversed biased.}\label{fig:RIS_Element}
\end{center}
\end{figure}

The general name for a reflective surface possessing subwavelength-size elements and intelligence to change its reflection properties is 
{\em metasurface\/} \cite{main-ris}.
To demonstrate the metasurface's capability of programmable reflection phase shifts allowing for the control and redirection of incident plane waves, we consider
an RIS made up of $M$ elements  along $x$
with unit cells as shown in Fig.~\ref{fig:RIS_Element}. Our nominal implementation of the RIS involves the use of square-shaped metal patches positioned on a grounded dielectric substrate. Varactors are placed at the center of each unit cell, connecting adjacent patches that are separated by gaps $w$. This design is a modified version of the dogbone-shaped metasurface discussed in \cite{Hanna22,Capolino13}, where the magnetic resonance effect enables the tunability of the reflection coefficient phase. The geometry of the design is shown in Fig.~\ref{fig:RIS_Element} and uses the substrate Rogers RT5880LZ as a dielectric spacer, with relative permittivity $\epsilon_{r} = 2$ and loss tan $\delta = 0.0021$, and  dimensions in mm given by $A = B = 19$, $A_1 = B_1 = 17.8$, $h = 1.27$, and $ w = 1.2$.

\begin{figure}[!t]
\begin{center}
  \includegraphics[width=1.8in]{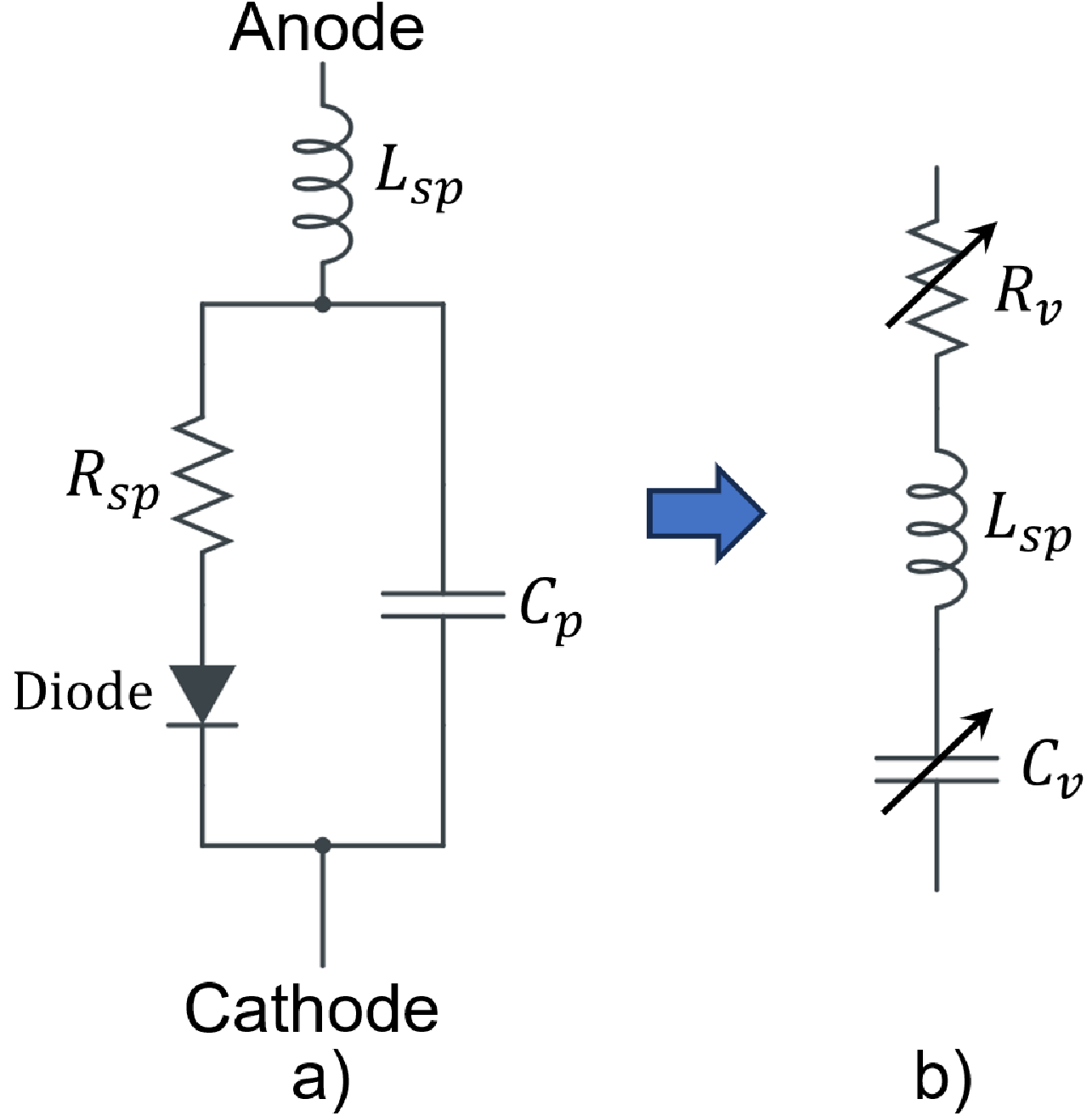}
  \caption{Circuit model of the varactor. (a) SPICE model provided by the vendor. (b) Simplified equivalent  RLC series ($R_v (V)$, $L_{sp}$, $C_v (V)$) circuit model. The values of $C_v$ and $R_v$ vary with the applied bias voltage.}\label{fig:Varactor_model}
\end{center}
\end{figure}

\begin{figure}[!t]
\begin{center}
  \includegraphics[width=2.8in]{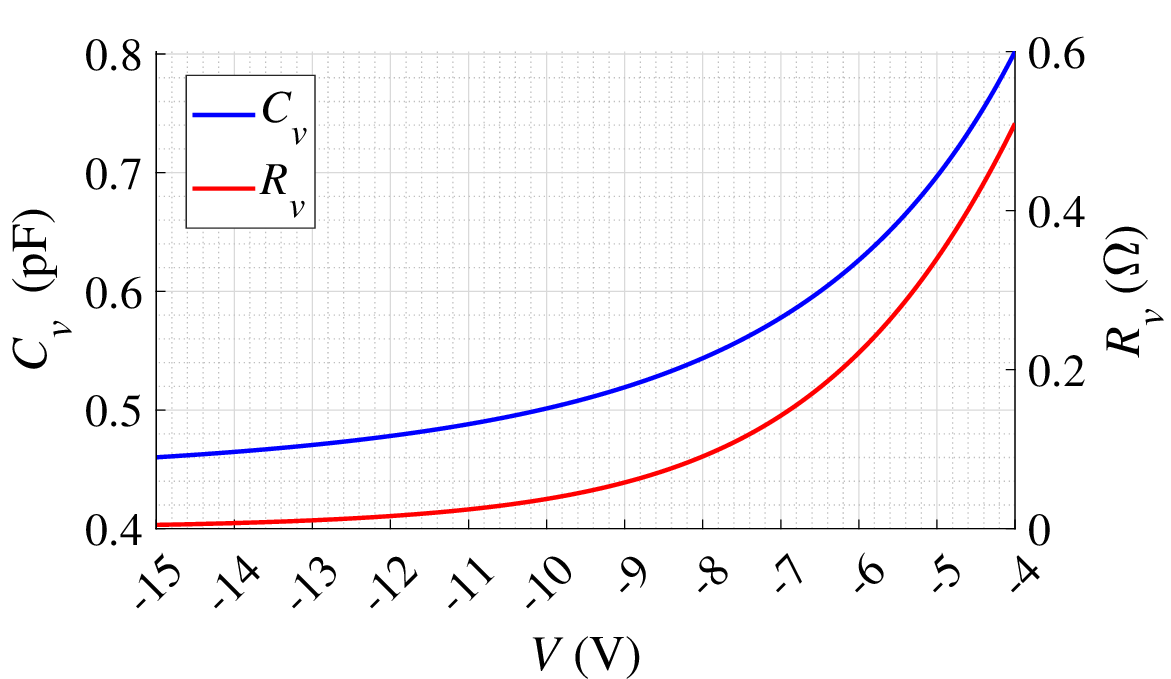}
  \caption{Equivalent capacitance and resistance of the varactor model in Fig. \ref{fig:Varactor_model} (b) as a function of the varactor biasing voltage. Knowledge of these two functions of $V$ leads to the analytic expression of the reflection coefficient $\phi_m(V)$ via (\ref{eq:ReflCoeff}), accounting for losses and RIS electromagnetic couplings.}\label{fig:Cv_Rv}
\end{center}
\end{figure}

To achieve reconfigurable behavior for each unit cell, we employ the SMV1231-040LF varactor, provided by Skyworks Solutions, Inc. This specific varactor is chosen due to its desirable characteristics, including a low series inductance $L_{sp}=0.45$ nH and resistance below 0.6 $\Omega$, which are important for the intended design. The nonlinear varactor model obtained from the datasheet is shown in Fig.~\ref{fig:Varactor_model}(a) 
and the small-signal model used in our equivalent RLC circuit model for the RIS is shown in Fig.~\ref{fig:Varactor_model}(b),
where, given varactor biasing voltage $V$, values for $R_v (V)$ and $C_v (V)$ are obtained from a parametric sweep simulation using Advanced Design System (ADS) software. In particular, the small-signal varactor impedance, $Z_{v}$, is computed from the S-parameter matrix of the model in Fig.~\ref{fig:Varactor_model}(a) for different reverse-bias voltages and the results are fit to match the impedance of the series RLC circuit.
In the simplified varactor model, the series inductance $L_{sp}$ is the package inductance and it is static, and the two additional elements are defined as $R_v (V) = \textrm{Re}\left(Z_{v}\right)$ and $C_v (V) = 1/\left(\omega^2L_{sp}-\omega\; \textrm{Im}\left(Z_{v}\right)\right)$.
The varactor capacitance tuning range is limited to 0.46 -- 0.8 pF, and the varactor resistance tuning range is limited to 0 -- 0.6 $\Omega$. These parameters are detailed in Table \ref{tab:Cv_Rv_Varactor} and illustrated in Fig.~\ref{fig:Cv_Rv}, where they are plotted as a function of the varactor biasing voltage.

\begin{table}[]
\caption{Values of the equivalent capacitance and resistance of the varactor model in Fig. \ref{fig:Varactor_model} (b) for different values of the varactor biasing voltage.}
\centering
\begin{tabular}{|l|l|l|}
\hline
\multicolumn{1}{|c|}{$V$ (V)} & \multicolumn{1}{c|}{$C_v$ (pF)} & \multicolumn{1}{c|}{$R_v$ ($\ohm$)} \\ \hline
-15 & 0.460 & 0.005 \\ \hline
-14 & 0.465 & 0.007 \\ \hline
-13 & 0.471 & 0.011 \\ \hline
-12 & 0.478 & 0.016 \\ \hline
-11 & 0.488 & 0.024 \\ \hline
-10 & 0.501 & 0.037 \\ \hline
-9  & 0.519 & 0.058 \\ \hline
-8  & 0.544 & 0.091 \\ \hline
-7  & 0.578 & 0.142 \\ \hline
-6  & 0.626 & 0.221 \\ \hline
-5  & 0.697 & 0.340 \\ \hline
-4  & 0.802 & 0.509 \\ \hline
\end{tabular}
\label{tab:Cv_Rv_Varactor}
\end{table}

\begin{figure}[!t]
\begin{center}
\noindent
  \includegraphics[width=1.6in]{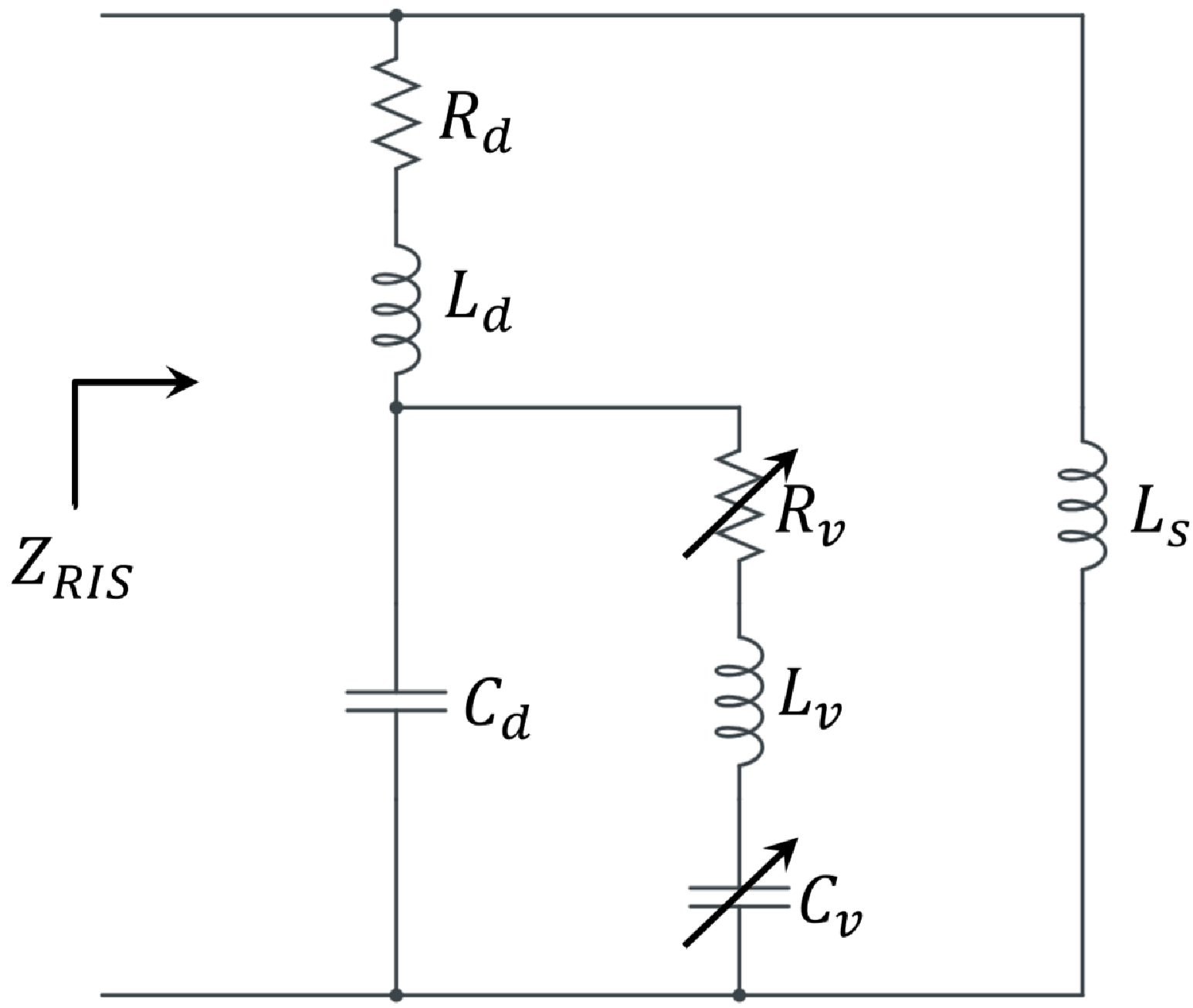}
  \caption{Equivalent analytical circuit model of the RIS. $Z_{RIS}$ is seen from the left.}\label{fig:circuit}
\end{center}
\end{figure}

A realistic RIS model is used to evaluate the reflection coefficient, as in \cite{Hanna22}, along the lines of \cite{Costa21}. The equivalent circuit model for plane wave reflection is shown in Fig.~\ref{fig:circuit}, where the parameters $R_d$, $C_d$, and $L_d$ are the resistance, capacitance, and inductance associated with the square-shaped unit cell element, and the inductance $L_s$ is an equivalent element that accounts for the grounded substrate, leading to the so called ``magnetic resonance" as explored in \cite{Capolino13} and also previously investigated in \cite{Sievenpiper99, Best08}.
The varactor is represented by the equivalent series RLC circuit model shown in Fig.~\ref{fig:Varactor_model}(b).

The RIS equivalent impedance, $Z_{eq}$, seen by a plane wave without considering the varactor is given by
\begin{equation}\label{eq:zin}
Z_{eq}= \left (R_d+j\omega L_d+ \frac{1}{j\omega C_d}\right )\, || \,j\omega L_s.
\end{equation}
This expression is rewritten as a function of the magnetic resonance, $\omega_m$ and electric resonance, $\omega_e$, as
\begin{equation}\label{eq:zin_wm_we}
Z_{eq}= \frac{j\omega L_{s}\left (1+j\omega R_{d}C_{d}-\left (\frac{\omega}{\omega_e}\right )^{2}\right )}{\left (1+j\omega R_{d}C_{d}-\left (\frac{\omega}{\omega_m}\right )^{2}\right )},
\end{equation}
where $\omega_e^2 = 1/\left (C_d\left (L_d+L_s\right )\right )$ and $\omega_m^2 = 1/\left (C_d L_d\right )$.
We note that close to (but not at) $\omega_m$ the reflection phase is 0 degrees, and at $\omega_e$, the reflection phase is almost 180 degrees, both studied in \cite{Capolino13}.
To acquire accurate numerical values for the elements $L_d$, $C_d$, and $R_d$, a single full-wave simulation without including the varactor is performed. The simulation models the RIS for plane wave orthogonal incidence by using a single cell with periodic boundary conditions, hence accounting for mutual couplings. It also accounts for dielectric and copper losses. The Z-parameters are evaluated from the S-parameters to obtain the values of $\omega_e$ and $\omega_m$. The inductance  $L_s = \mu_0 h = 1.6$ nH is analytically determined by modeling the substrate as a short-circuited transmission line section with a length of $h$ and approximating the expression of the impedance as $Z_s = j\omega \tan\left(\mu_0h\right) \approx j\omega \mu_0h$. The other values are obtained as $L_d = L_s/\left(\left(\omega_e/\omega_m\right)^2-1\right) = 0.39$ nH, $C_d = 1/\left(L_d \omega_e^2\right) = 0.53$ pF, and $R_d = 
L_s/\left(C_d\left(1+L_d/L_s\right)\textrm{Re}\left(Z_{eq}\left(\omega_m\right)\right)\right) = 0.08$ $\ohm$.

The varactor included in the analytical model 
is in parallel 
to the capacitor $C_d$ that models the capacitance created by the gap across which the varactor is connected.
Note that the inductance $L_v = L_{sp} + L_p = 2.34$ nH replaces the inductance $L_{sp}$. The term $L_p$ represents the parasitic inductance introduced by the varactor when connected across the gap in the full-wave simulations, which will be presented later.
Therefore, the total equivalent RIS impedance, $Z_{RIS}$, is given by
\begin{equation}\label{eq:zRIS}
\begin{split}
& Z_{RIS} = \\
& \left (R_d+j\omega L_d+ \left (R_v+j\omega L_v+ \frac{1}{j\omega C_v}\right )||\frac{1}{j\omega C_d}\right )||j\omega L_s,
\end{split}
\end{equation}
and the reflection coefficient, $\phi$, is evaluated as
\begin{equation}
\phi= \frac{Z_{RIS}-Z_0}{Z_{RIS}+Z_0},
\label{eq:ReflCoeff}
\end{equation}
where $Z_0$ is the free-space impedance.

\begin{figure}[!t]
\begin{center}
  \includegraphics[width=2.8in]{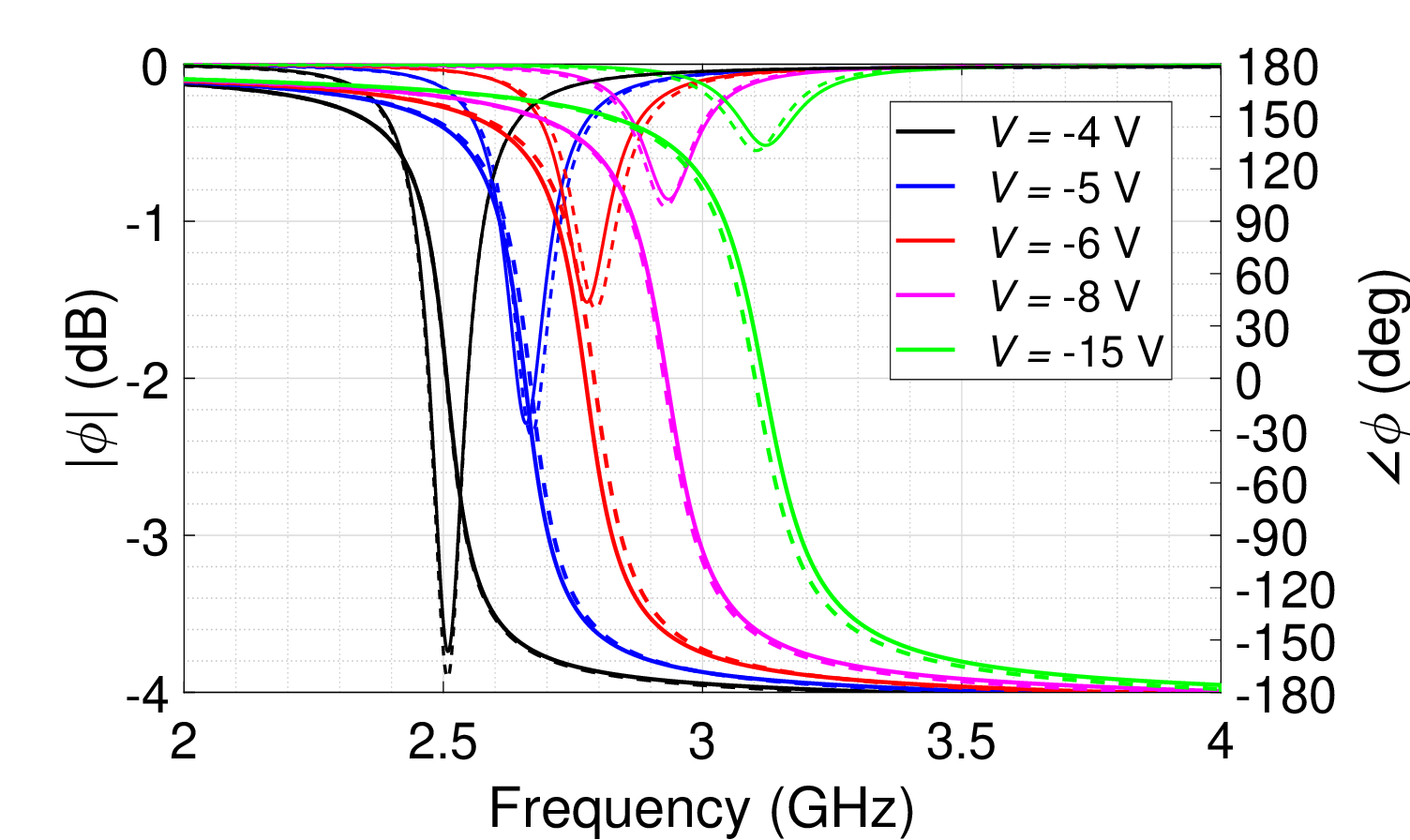}
  \caption{Magnitude and phase of the RIS reflection coefficient varying varactor's biasing voltage, calculated using the equivalent circuital analytical model (solid lines), and compared with the results of the full-wave simulation (dashed lines). The analytical model accounts for metallic and substrate losses as well as losses in the varactors. It also accounts for electromagnetic couplings among the RIS elements, calculated based on the local periodicity approximation. Model and full-wave simulations are in good agreement.}\label{fig:Ref_Coef_M_P}
\end{center}
\end{figure}

\begin{figure}[!t]
\begin{center}
  \includegraphics[width=2.8in]{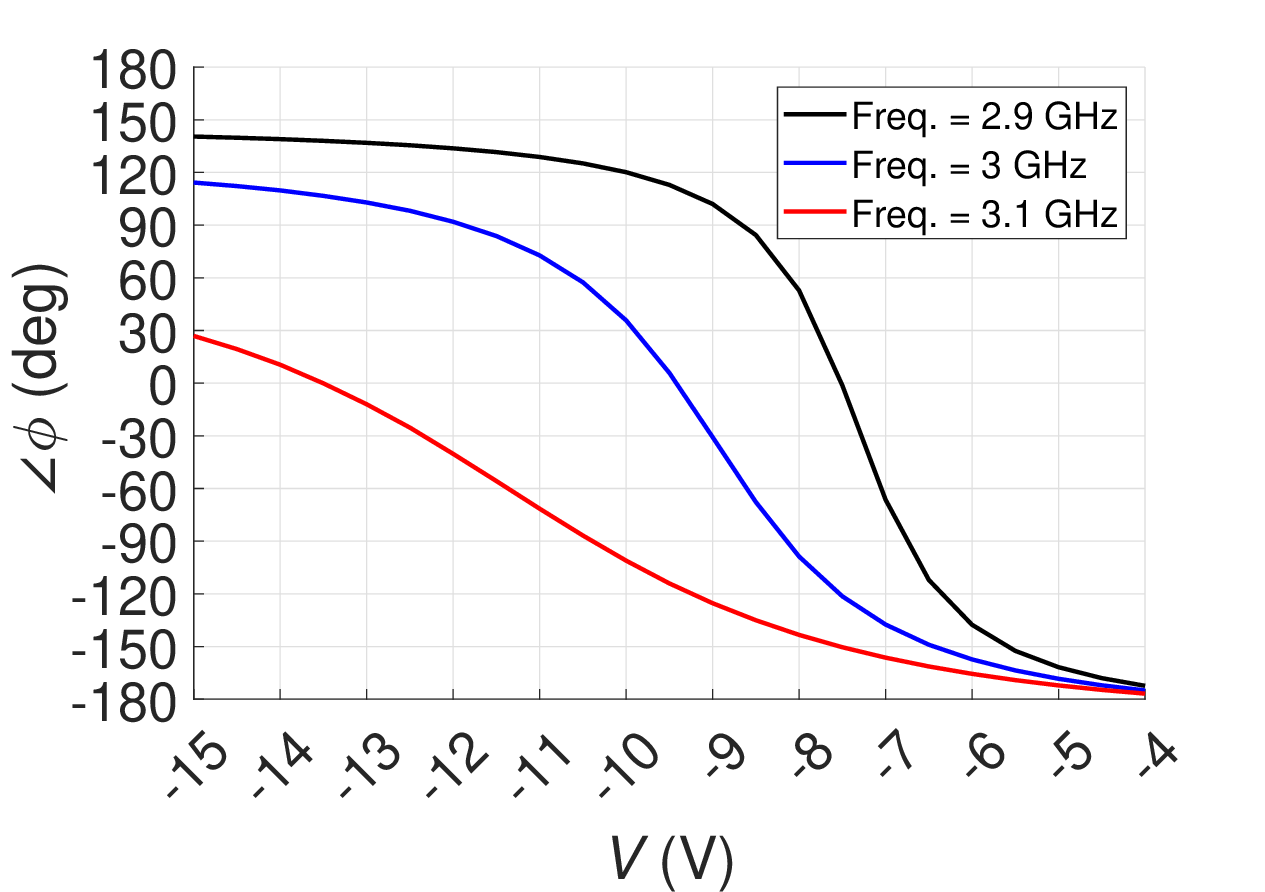}
  \caption{Phase of the RIS reflection coefficient as a function of the biasing voltage applied to the varactor for three different frequencies.}\label{fig:Phase_Ref_Coef}
\end{center}
\end{figure}

In order to assess the performance of the proposed analytical model, the commercial CST Studio Suite software package is used to obtain the reflection coefficient from full-wave simulations, including the effect of the varactor as a lumped load. The magnitude and phase of the reflection coefficient for various varactor reverse bias voltages are plotted in Fig.~\ref{fig:Ref_Coef_M_P}, demonstrating the capability of the circuit model to estimate the reflection coefficients for various frequencies and varactor voltages.
The results demonstrate that a phase dynamic range (defined as the set of phase values that can be obtained at a given frequency) of around $290^\circ$ is activated
in the band between 2.6 GHz--3 GHz. The phase of the RIS reflection coefficient as a function of the biasing voltage applied to the varactor for three different frequencies is shown in  Fig.~\ref{fig:Phase_Ref_Coef}, where a tradeoff between the phase dynamic range and the biasing voltage range can be observed.


\section{Wave-Controlled RIS (Full-Domain Control Basis)}
%
We show that individual control on the biasing voltage is achieved using a superposition of full-domain functions,  $w(x,t) = \sum_n w_n(x,t)$, over the whole RIS length as shown in Fig. \ref{fig:RIS_Figure_1}. In particular, we use a set of $N$ standing waves over the whole length of the RIS, written as
\begin{equation}
    w(x,t) = W_0 + \sum_{n=1}^{N} W_n  \sin\left(\frac{n \pi (x+L_{\it left})}{L_{\it tot}}\right) \sin (n \omega_b t ) ,\label{eqn:Wxt}
\end{equation}
where $N$ represents the number of full-domain expansion modes  in the bias voltage decomposition and $W_n$ is the amplitude of the $n$-th mode, $n =0, 1, \ldots, N$. The vector of coefficients ${\bf W}=[W_0, W_1, \ldots ,W_{N}]^T$ is used to parameterize the biasing voltage. We consider $L_{\it tot}=L+L_{\it left}+L_{\it right}$ because the two extra segments on the left and right of the biasing TL are useful to better control the voltage values on the RIS over the length $L$.
Note that (\ref{eqn:Wxt}) corresponds to a {\em truncated\/} Fourier series in space, with $N$, rather than an infinite number of sinusoids. What is shown in (\ref{eqn:Wxt}) is a signal that will be generated in the biasing TL for control of the RIS that, when sampled, will yield the needed bias voltage at each RIS element. For this reason,
the value of $N$ is desired to be as small as possible to limit the variation in $w(x,t)$ with $x$, and also to reduce the control signaling overhead.  The biasing voltage is sampled along the biasing TL and applied as inputs to the RIS elements' varactors at positions $x_m=m d_x, m=0,1,\ldots,M-1$, where $d_x$ is the distance between the centers of each pair of adjacent RIS elements.

In the development of the biasing TL, it is convenient to use low frequencies for the standing waves, much smaller than the RIS operation frequency that is either in the cm-wave (i.e., microwave) or in the mm-wave range. This is achieved by considering a slowness factor $n_{slow}$ of the waves in the biasing TL that is dependent on the materials used and the actual geometry of the biasing TL. Therefore, in  the biasing TL, the phase velocity of the waves along the $x$ direction is equal to $v_{ph}=c/n_{slow}$, where $c$ is the speed of light. The fundamental standing wave depicted by $w_1(x,t)$ in Fig.~\ref{fig:RIS_Figure_1}  is such that $k_b L_{\it tot} = \pi$, where $k_b=\omega_b/v_{ph}$ is the wavenumber, $\omega_b$ is the angular frequency, and $L_{tot}$ is the total length of the biasing TL in the $x$ direction. Therefore, the fundamental standing wave oscillates at  $f_b=\omega_b/(2\pi)$ where $\omega_b= \pi v_{ph}/  L_{\it tot}$  \cite{MKF-ris}.  A simple choice of parameters can produce a value for $f_b$ in the low MHz range. Higher order standing waves $w_n(x,t)$, $n=2,3, \ldots,N$, oscillate at frequencies $n f_b$,
with wavenumbers $k_{b,n}=n k_b$, $n=1,2,\ldots,N$.
In this model, we assume that $0\le x_m \le L$, $m = 0, 1, \ldots, M -1$\footnote{In reality, the TLs underneath the RIS surface have different lengths between the RIS elements than on the RIS surface. In our experimental implementation, we use a serpentine structure for TL. This is done to reduce spatial sensitivity in realizing the TLs. Let us say the length of the TL between two adjacent RIS elements is $d_x$. Then $L = (M-1) d_x$, $L_{\it left}=M_l d_x$, $L_{\it right}=M_r d_x$ where $M_l$ and $M_r$ are nonnegative numbers, and $L_{tot} = L + L_{\it left} + L_{\it right}$.
}.
For notational convenience, the standing waves in (\ref{eqn:Wxt}) are rewritten directly in terms of $m$ as
\begin{equation}
\begin{split}
    w & (md_x,t) = \\
    & W_0 + \sum_{n=1}^{N}{W_n \sin\left(\frac{n \pi (m+M_l)}{M-1+M_l+M_r}\right)\sin(n\omega_b t)},
    \label{eqn:wmt1}
    \end{split}
\end{equation}
where $M_l=L_{\it left}/d_x$ and $M_r=L_{\it right}/d_x$.
\subsection{Envelope Detector Circuit}
\begin{figure}[!t]
\centering
\includegraphics[width=0.3\textwidth]{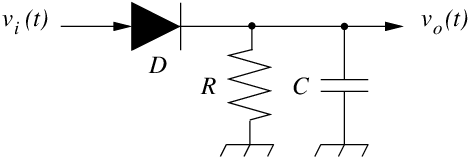}
\caption{Rectifier circuit used to rectify the alternating current voltage on the biasing TL. The input voltage $v_i(t)$ is the standing waves at location $m$; the output voltage $v_o(t)$ is the rectified voltage to bias the varactor at location $m$. The circuit follows the envelope or the peak of $v_i(t)$ via the diode $D$, the resistor $R$, and the capacitor $C$. The time constant $RC$ should be chosen sufficiently large to keep the capacitor discharge to manageable levels so that $v_o(t)$ does not show a significant drop between the consecutive peaks of $v_i(t)$. (This circuit is not present when the sample-and-hold technique is used, as described later on.)}
\label{fig:envelopedetector}
\end{figure}

A potential way to detect the voltage level needed for biasing the varactors is by using the rectifier circuit shown in Fig.~\ref{fig:envelopedetector}, one per RIS element. This is a conventional circuit element employed in communications electronics, most commonly to demodulate an amplitude-modulated continuous-time signal. Its operating principles are simple, see, e.g., \cite{Haykin94}. Typically, the time constant $RC$ is chosen such that
\begin{equation}
\frac{1}{f_N} \ll RC \ll \frac{1}{f_{\it reconfig}}
\end{equation}
where $f_N$ is the highest frequency of the sinusoidal signal in the biasing TL (i.e., related to the highest $n$-harmonic). The value for $f_{\it reconfig}$ is the frequency at which the RIS needs to be reconfigured. This condition ensures the circuit is able to follow the envelope of the highest-frequency sinusoid in the biasing TL. We employ the envelope detector to perform a peak detection of the standing wave signal which oscillates with time $t$.
The standing waves on the biasing transmission line of the RIS oscillate in time with 
frequencies $n f_b$ as $\sin(n\omega_b t)$, hence the highest frequency is $f_N=Nf_b$.

To only sample the peak of the standing wave at each element $m$ over time, the rectifier outputs are simply described here by taking the peak (maximum) values of the alternating time domain signal, as
\begin{equation}
    \begin{split}
    & w(m) = \\
    & \max_t{\left(W_0 + \sum_{n=1}^{N}{W_n \sin\left(\frac{n \pi (m+M_l)}{M-1+M_l+M_r}\right)\sin(n\omega_b t)}\right)},
    \end{split}
\label{eqn:w_m-max}
\end{equation}
where $w(m)$ represents the DC voltage bias supplied to each varactor index using the standing waves.
We observe that the envelope of the time-varying part inside the parenthesis in (\ref{eqn:w_m-max}) is symmetric in its positive and negative ranges. Since varactors are polarized inversely, we have decided to work with the negative part of the envelope and thus in the sequel, we will replace $\max$ in (\ref{eqn:w_m-max}) with $\min$. In addition, since the DC level of the standing wave is independent of time, the expression is simplified to
\begin{equation}
\begin{split}
    & w(m) = \\
    & W_0+\min_t{\left(\sum_{n=1}^{N}{W_n \sin\left(\frac{n \pi (m+M_l)}{M-1+M_l+M_r}\right)\sin(n\omega_b t)}\right)}.
    \label{eq:envelope_detector_wave_model}
\end{split}
\end{equation}
Due to the $\min$ function, using rectifiers implies a nonlinear relationship between the standing wave coefficients $W_n$ and the spatial voltage levels $w(m)$.
\subsection{Sample-and-Hold Circuit}
\begin{figure}[!t]
\centering
\includegraphics[width=0.3\textwidth]{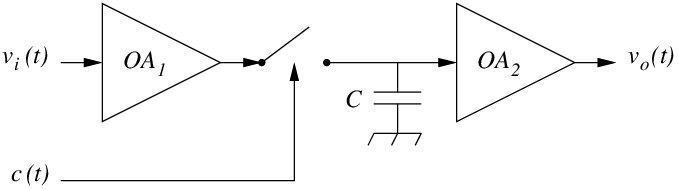}
\caption{Sample-and-hold circuit to bias the varactors. The input voltage $v_i(t)$ is the biasing standing waves at location $m$; the output voltage $v_o(t)$ is used to bias the varactor at location $m$, and the control signal is $c(t)$. ${\it OA}_1$ and ${\it OA}_2$ are operational amplifiers, and $C$ is the capacitance that holds the sampled voltage. (This circuit is not present when we use the rectifiers.)}
\label{fig:sample-hold}
\end{figure}

Another potential way to detect the voltage level needed for biasing the varactors is by using sample-and-hold (SH) circuits shown in Fig. \ref{fig:sample-hold}.  In this approach, every RIS element employs an SH circuit to sample the standing wave along the transmission line and hold it for a given duration to configure the corresponding RIS element. The  SH is a standard circuit element used in many applications, for example, in analog-to-digital converters \cite{HH80}. A conceptual diagram is provided in Fig.~\ref{fig:sample-hold} where the input voltage $v_i(t)\equiv w(md_x,t)$ is sampled at the output of the operational amplifier ${\it OA}_1$ under the control of the signal $c(t)$. This signal is held in the capacitor $C$ such that it can be read out at the output of the operational amplifier ${\it OA}_2$ as the bias voltage for the varactor diode controlling the phase of the RIS element.
These circuits are used in analog-to-digital converters to eliminate variations in an input signal because such variations can corrupt the conversion. As shown in Fig.~\ref{fig:sample-hold}, a sample-and-hold circuit has a switching device such as a transistor which loads the capacitor $C$ with the sampled voltage. This happens during the {\em sample\/} stage of the circuit when the buffer amplifier ${\it OA}_1$ charges or discharges the capacitor and makes the voltage across $C$ equal to the sampled input voltage. In the next stage, the {\em hold\/} stage, the switch disconnects the capacitor from ${\it OA}_1$, which can be read out by ${\it OA}_2$. It is possible that the capacitor can discharge through the load it sees at the input of ${\it OA}_2$ and its own leakage, but this can be made to take a long time.
\subsubsection{Distribution of the Sampling Signal}
%
Only a single control signal is required for the SH circuits at each element, since they can be sampled at or near the same time. A coaxial cable connection can be used to eliminate interference between the control and standing wave signals.
The sampling signal requires less bandwidth than the standing wave.
This simple configuration assumes all RIS elements are provided with the same sample timing. Even if different sampling times are used at different RIS elements, it is possible to orthogonalize the signal. The coaxial cable will prevent interference as long as the cut-off frequency is not approached.

An alternative for distribution of the sampling signal is wireless transmission. In such a system, the wireless module would be connected to the control inputs of the analog switches responsible for the operation.
%
\subsubsection{Distribution of Power}
The power can be distributed by a single-wire DC distribution circuit. It is possible to carry out this power distribution such that the possibility of RF interference can be avoided. In fact, the same coaxial cable for distribution of the sampling signal can also carry the DC power. It is possible that some sample-and-hold circuits would require more than one voltage level. In that case, use of more than one cable is possible, or multiple DC voltages can be derived from a single voltage source. We note that the sample-and-hold circuits are in general not power hungry, and therefore, distribution of power will not require a substantial effort.

As an alternative, power can be locally generated at each RIS element by means of energy harvesting. For example, energy can be harvested from light and stored at night. Or energy can be harvested from received RF energy. Yet another alternative is to use batteries with replacement; for example, one can alternate between two batteries for hitless operation. Reference \cite{VKPSVB23} discusses the use of RFID tags to power the entire RIS.
\subsubsection{Design of Sample-and-Hold Circuits}
A number of criteria need to be judiciously applied to the design of a sample-and-hold circuit. 
Examples are switching speed, settling times, aperture time, jitter and noise, input range, power consumption, etc.
%
%
\section{Optimization Algorithms}
In the communication theory literature, algorithm design for RIS optimization has almost exclusively employed simplistic models in which one has the ability to directly and independently control the reflection coefficient $\phi(m)$ of each RIS element. In reality, the actual control signal at the $m$-th unit cell is (for example) a biasing voltage $V(m)$ on a varactor diode, and as shown in the realistic unit cell model presented earlier, arbitrarily tuning the phase of $\phi(m)$ is not possible. Furthermore, in the approach considered here, $V(m)$ is obtained by sampling a set of standing waves $w(md_x,t)$ using a device that is neither linear nor time variant. As a result, compared with conventional RIS optimization methods, it is significantly more challenging to design the weights $W_n$ to produce a standing wave $w(md_x,t)$ that when sampled yields a voltage $V(m)$ that in turn generates the desired RIS response $\phi(m)$. Achieving this goal requires approaches entirely different from those proposed to date in the literature which only consider optimization of $\phibf$ directly. In this section we present the results of several algorithms for solving this problem that differ based on the desired performance metric and the type of sampling circuit used to extract the varactor biasing. We focus on scenarios where the RIS is designed to form beams or nulls in certain directions in response to a line-of-sight signal from a transmitter.

\subsection{Mathematical Formulation of the Optimization Problem}
%

For the purpose of describing the optimization of the proposed RIS architecture, we assume that the direct propagation path between the transmitter and each UE is either already known or assumed to be nonexistent, and we only consider the path reflected by the RIS. Let the narrowband flat fading channel between the single-antenna transmitter and RIS element $m$ be described by $g(m)$ and that between RIS element $m$ and the $k$-th single-antenna UE be $h_{k}(m)$. We assume perfect knowledge of $h_{k}(m)$ and $g(m)$ for all $K$ receivers and all $M$ RIS elements. Then, the expression for the signal received by UE $k$ is
\begin{equation}
    y_k=\left[\sum_{m=0}^{M-1}{h_{k}(m) \phi(m) g(m)}\right]s_k+n_k \; ,
\end{equation}
where $\phi(m)$ is the reflection coefficient at the $m$-th RIS element, $s_k$ is the transmitted signal, and $n_k$ is additive white Gaussian noise (AWGN), {i.e.,} $n_k\sim \mathcal{CN}(0,\sigma_s^2)$. Writing this in matrix form, we have
\begin{equation}
    y_k=\boldsymbol{h}_k^T\boldsymbol{\Phi}\boldsymbol{g}s_k+n_k
\end{equation}
where $\boldsymbol{h}_k=[h_{k}(0),h_{k}(1),\ldots,h_{k}(M-1)]^T$ and $\boldsymbol{g}=[g(0),g(1),\ldots,g(M-1)]^T$ are respectively the $M\times1$ channels from the RIS to UE $k$ and the BS to the RIS, and the RIS response is described by the diagonal matrix $\boldsymbol{\Phi}=\diag[\phi(0),\phi(1),\ldots,\phi(M-1)]$. Each $h_{k}(m)=\alpha_{k}(m)e^{-j\theta_{k}(m)}$ and $g(m)=\beta(m)e^{-j\psi(m)}$. Since the RIS elements are passive (their reflection coefficients are only determined from the capacitance supplied by the varactors), $|\phi(m)|\leq1$ for all $m=0,1,\ldots,M-1$.

The signal-to-noise ratio (SNR) at UE $k$ is the ratio of the received signal power divided by the noise power $\sigma_s^2$:
\begin{equation}
    \snr_k=\frac{|E[y_k]|^2}{\sigma_s^2}=\frac{|E[\boldsymbol{h_k^T \Phi g}x]|^2}{\sigma_s^2}=\frac{\rho_s|\boldsymbol{h_k^T\Phi g}|^2}{\sigma_s^2}
    \label{eq:SNR}
\end{equation}
where $\rho_s$ is the average power for each transmitted symbol.
To evaluate the ability of the proposed wave-controlled RIS to configure the wireless channels to the UEs, we will consider optimizing the RIS configuration $\boldsymbol{\Phi}$ for the two objective functions described below.
\begin{enumerate}
    \item Maximize the SNR for a given UE:
    \begin{equation}
        \max_{\boldsymbol{\eta}}{\snr}=\max_{\boldsymbol{\eta}}{\rho_s|\boldsymbol{h}_k^T\boldsymbol{\Phi g}|^2} \; ,
        \label{eq:SNR_max}
    \end{equation}
    where $\boldsymbol{\eta}$ is a parameter vector that represents the variables that control the RIS configuration $\boldsymbol{\Phi}$. These variables can be the reflection coefficients themselves
    ($\boldsymbol{\eta}=\boldsymbol{\phi}=[\phi(0),\phi(1),\ldots,\phi(M-1)]^T$), the biasing voltages at the varactors ($\boldsymbol{\eta=V}=[V(0),V(1),\ldots,V(M-1)]^T$), or amplitudes of the modes that define the
    biasing waveforms ($\boldsymbol{\eta=W}=[W_0,W_1,\ldots,W_N]^T$).
    \item Maximize the worst-case signal-to-leakage-plus-noise ratio (SLNR) for a certain combination of desired and undesired receivers:
    \begin{equation}
        \max_{\boldsymbol{\eta}}{\slnr}=\max_{\boldsymbol{\eta}}{\frac{\min_{i\in\{1,2,\ldots,K\}}{\rho_s|\boldsymbol{h}_{d,i}^T\boldsymbol{\Phi g}|^2}}{\max_{j\in\{{1,2,\ldots,L\}}}{\rho_s|\boldsymbol{h}_{e,j}^T\boldsymbol{\Phi g}|^2}+\sigma_s^2}}.
        \label{eq:SLNR_max}
    \end{equation}
    where $\boldsymbol{h}_{d,i}$ are the channels between each RIS element and each desired Rx, and
    $\boldsymbol{h}_{e,j}$ are the channels corresponding to undesired or ``eavesdropping'' receivers. The worst-case SLNR is calculated using the minimum power reflected towards any of the $K$ desired receivers, divided by the summation of the noise power and the maximum power reflected towards any of the $L$ undesired receivers.
\end{enumerate}


In order for the analytical model to match the physical RIS model created from our full-wave simulations, the following assumptions are made for the numerical examples \cite{MKF-ris}
\begin{itemize}
    \item The RIS is arranged as a uniform linear array with elements separated by $\Delta,$ which is in terms of wavelengths, and therefore a unitless quantity.
    \item The BS is located in the far field in the direction of the broadside of the RIS, such that there is normal incidence between the BS and each RIS element, and hence
\[
\boldsymbol{g} =
          [1, 1, \ldots, 1]^T.
\]
    \item The UE is located at an azimuth angle of $\theta^*$ from the RIS and hence
\[
\begin{split}\nonumber
 \boldsymbol{h}(\theta^*) & =  \\
&    \hspace{-1em}\left[\begin{array}{llllll}1,&\hspace{-0.5em}e^{-j\kappa(\theta^*)},&\hspace{-0.5em}e^{-j2\kappa(\theta^*)},&\hspace{-0.5em}\ldots,&\hspace{-0.5em}e^{-j(M-1)\kappa(\theta^*)}\end{array}\right]^T
\end{split}
\]

    where $\kappa(\theta)=2\pi\Delta\sin(\theta)$. We will assume a specific case with a carrier frequency of $f_c=3$ GHz and a spacing of 19 mm between RIS elements. At $f_c=3$ GHz, this corresponds to a $\Delta$ of about 1/5 = 0.2. $\boldsymbol{h}(\theta^*)$ assumes line-of-sight channels to the users, though this is not strictly necessary.
\end{itemize}
\subsection{Optimization Problem 1: Maximizing SNR at a Single UE Direction}
\subsubsection{Ideal Phase}
In this case, assume that each individual reflection coefficient $\phi(m)$ can be modified to any value such that $|\phi(m)|=1$, so there is full control over the phase shift of the reflected wave. The task is to find $\boldsymbol{\Phi}$ such that the expression \(P=\rho_s|\boldsymbol{h}^T \boldsymbol{\Phi g}|^2\) is maximized.

For this analysis, the average transmission power $\rho_s$ can be ignored since the RIS has no impact on $\rho_s$ and does not provide any amplification or attenuation due to its reflective nature. For the flat fading model $h(m)=\alpha(m) e^{-j\theta(m)}$ and $g(m)=\beta(m) e^{-j\psi(m)}$, the power is maximized when $\phi(m)=e^{j(\theta(m)+\psi(m))}$, since this produces a coherent sum:
\begin{equation}
\begin{split}
    P & =\rho_s\left|\sum_{m=0}^{M-1}{h(m)\phi(m) g(m)}\right|^2 \\
& =\rho_s\left|\sum_{m=0}^{M-1}{\alpha(m) e^{-j\theta(m)} e^{j(\theta(m)+\psi(m))} \beta(m) e^{-j\psi(m)}}\right|^2\\
& =\rho_s\left|\sum_{m=0}^{M-1}{\alpha(m) \beta(m)}\right|^2 . \label{eq:ideal_phase}
\end{split}
\end{equation}
\subsubsection{Arbitrary Voltage Bias}\label{sec:AVB}
Since the varactor diodes have voltage limits, some of the phase values may not be achievable, as seen in \frenchspacing Fig.~\ref{fig:Ref_Coef_M_P}--Fig.~\ref{fig:Phase_Ref_Coef}\nonfrenchspacing. Taking this constraint into account, the objective should be to set the bias voltage such that the resulting phase is as close as possible to the ideal phase of the previous case. We see from
Fig.~\ref{fig:Phase_Ref_Coef} that for frequencies in the range 2.9-3.1 GHz, the mapping from the varactor voltage to the reflection coefficient phase is a one-to-one function.
Let $\varphi(V(m))$ be the one-to-one mapping that converts the varactor voltage $V(m)$ into RIS reflection coefficient phase $\phase{\phi(m)}$ for RIS element $m$. Then, the phases are bounded
between $\phase{\phi_{min}}=\varphi(V_{max})$ and $\phase{\phi_{max}}=\varphi(V_{min})$. Therefore, the phases obtained by the biasing voltages become
\begin{equation}
    \phase{\phi_{arb}}=
    \begin{cases}
        \phase{\phi_{min}} & \mbox{if $\phase{\phi_{ideal}} < \phase{\phi_{min}}$},\\
        \phase{\phi_{ideal}} & \mbox{if $\phase{\phi_{min}} \leq \phase{\phi_{ideal}} \leq \phase{\phi_{max}}$} ,\\
        \phase{\phi_{max}} & \mbox{if $\phase{\phi_{ideal}} > \phase{\phi_{max}} .$}
    \end{cases}
\end{equation}
The ``arbitrary voltages" that reproduce these phase shifts are
\begin{equation}
    V=\varphi^{-1}(\phase{\phi_{arb}}) .
    \label{eq:phase_to_voltage}
\end{equation}
For the simulations to be presented later, the inverse mapping $\varphi^{-1}(\cdot)$ is obtained by linearly interpolating the phase values obtained for a set of discrete biasing voltages spaced with steps of 5mV. For the model described in Section~\ref{sec:model}, the biasing voltage range is $[-15V, -4V]$. The voltage values can interchangeably be represented by positive numbers for the same range of absolute values, assuming that the varactor diodes are inversely biased.
\subsection{Solution for Optimization Problem 1: Envelope Detector}
Recall that in the envelope detector model, only the peak of the standing wave voltage $w(x_m,t)$ at each varactor is converted into a sampled DC voltage. For negative biasing voltages, the most negative voltage is considered to be the peak (minimum). Therefore, the voltage at each varactor is modeled as in \eqref{eq:envelope_detector_wave_model} which tracks the negative signs.
\begin{figure}[!t]
    \centering
    \includegraphics[width=1.0\linewidth]{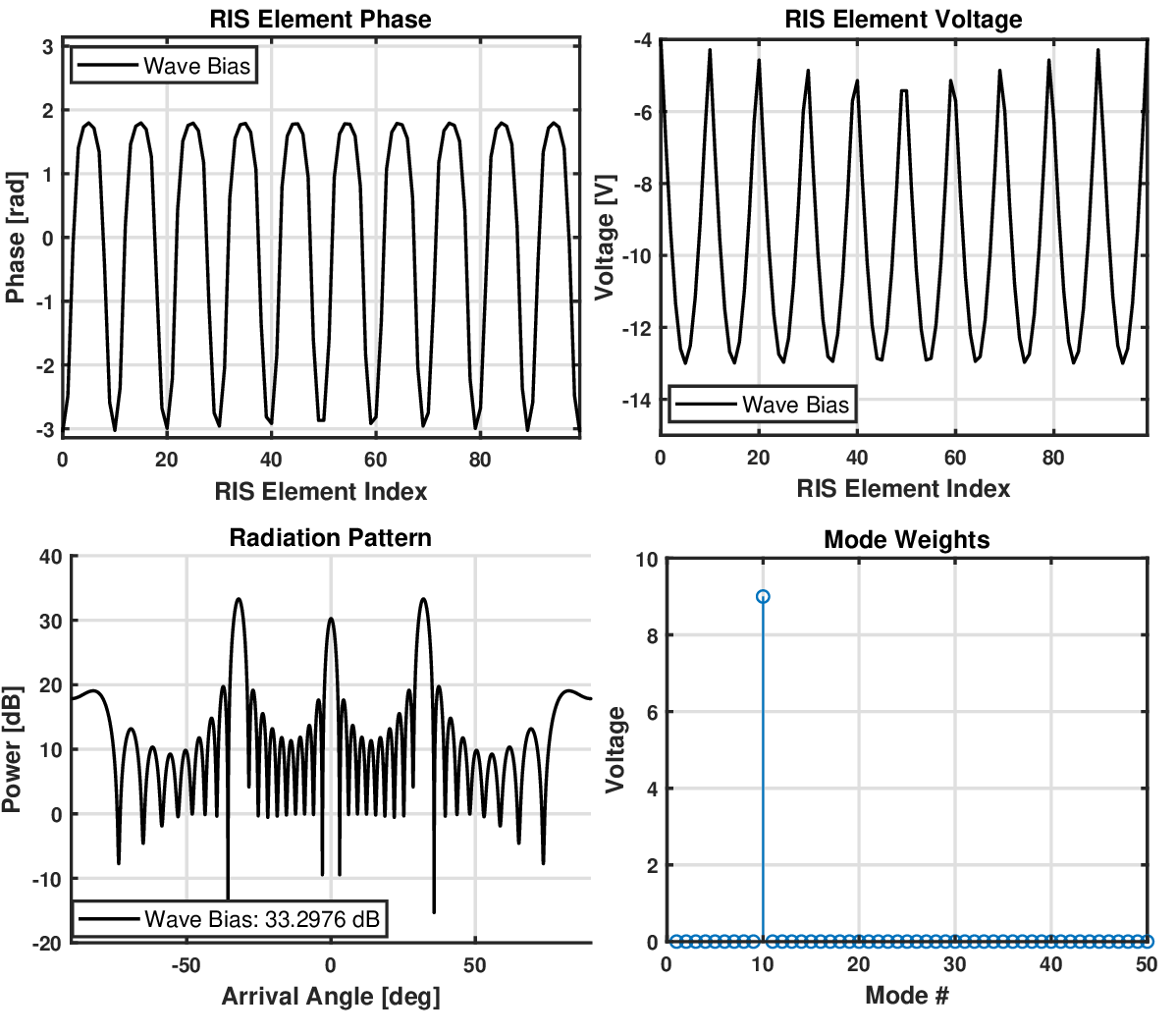}
    \caption{Envelope detector model with only one mode activated. Bottom right: $W_{10}=9$V, while the other mode amplitudes are zero. Top right: DC voltages across all 100 varactors, after rectification. Top left: Reflection phases created at each element. Bottom left: Resulting radiation pattern for a wide range of arrival angles.}
    \label{fig:RIS_Single_Mode}
\end{figure}

The problem of finding the weights to produce a desired peak voltage is a difficult nonlinear optimization problem that cannot be solved analytically. To cover the full range of the varactor biasing voltages, $\min_t(\cdot)$ will always produce a negative voltage, so $W_0$ should be set to the maximum voltage level (in this case, -4V). Of the multiple approaches that we have explored to maximize the power towards a single receiver, a combination of the two algorithms described below has proven to be fruitful.
\subsubsection{Algorithm 1 - Weight Ranking}
The motivation for this approach began with the observation that increasing only a single mode amplitude can greatly enhance power reflected towards a given direction, though changing one mode would not achieve the global solution since all the elements must be utilized accurately, requiring more dimensions for the optimization by wave biasing. It was noted that
there was a correlation between the frequency of the mode whose amplitude was increased and the direction towards which the RIS reflected the signal, as seen in Fig.~\ref{fig:RIS_Single_Mode}. Just by increasing the amplitude for mode \#10 ($W_{10}$) to 9V, without any contribution from the other modes, a gain of around 33.3 dB at both 32\degree~and -32\degree~was observed. Based on this, we developed the {\em Weight Ranking} algorithm, which ranks the weights by their importance to maximize power reflected towards a single desired receiver direction $\theta^*$. The algorithm is implemented as specified in Algorithm~1.

\begin{algorithm}[!t]
    \caption{Weight Ranking}\label{alg:weight_ranking}
    \begin{algorithmic}[1]
        \State $\boldsymbol{W} \gets [0,0,\ldots,0]^T$ (array containing $N$ amplitudes)
        \State $W_0 \gets -4$V
        \State $P_{arr} \gets [0,0,\ldots,0]$ (array containing $N$ entries of power measurements at the desired receiver angle)
        \For {each $n\in \{1,...,N\}$}
            \State Find the value for the weight $\boldsymbol{W}(n)$ that maximizes the power at $\theta^*$ by either increasing or decreasing the amplitude by 0.001 and calculating the voltage curve and power gain.
            \State Record the maximal power in $P_{arr}(n)$
            \State Reset $\boldsymbol{W}=[0,0,\ldots,0]^T$
        \EndFor
        \State Sort $P_{arr}$ in descending order and extract the indices $n$ that correspond to power values from highest to lowest that can be mapped later to corresponding amplitudes $\boldsymbol{W}(n)$.
    \end{algorithmic}
\end{algorithm}
\subsubsection{Algorithm 2 - Brute Force Optimization}
This algorithm takes the indices that correspond to the most influential amplitudes as determined in Algorithm~1, and uses a hill-climber approach to converge towards optimal amplitude weights by increasing or decreasing
each weight according to the order that was previously obtained \cite{hill-climber}. The weights are ordered before using the hill-climber approach since the order of the amplitudes matters when optimizing.
First, there is the constraint that the voltage must stay within the range $[-15V, -4V]$ so the summation of the weights is also constrained. Second, increasing the weights in a different order may cause the waveform
to change so that the contributions of the corresponding modes to the radiation pattern changes, due to the nonlinear relationship created by the $\min_t(\cdot)$ operation.
\begin{algorithm}[!t]
    \caption{Brute Force}\label{alg:brute_force}
    \begin{algorithmic}[1]
        \State $\boldsymbol{W} \gets [0,0,\ldots,0]^T$
        \State $W_0 \gets -4$
        \State calculate $w(m)$ using~(\ref{eq:envelope_detector_wave_model}), $m=0,1,...,M-1$
        \State $P_{old} \gets $ initial power reflected towards $\theta^*$ using $w(m)$
        \State $\mu \gets 1.0$ (initial step size)
        \State ``negate" $\gets$ 0 (Boolean value that determines if the current step is positive or negative)
        \Repeat
            \For {each $\boldsymbol{W}(n)$, starting from the highest power index to the lowest power index obtained from Algorithm \ref{alg:weight_ranking} }
                \State $\boldsymbol{W}_{new}(n) \gets \boldsymbol{W}(n)+\mu$
                \State Calculate $w_{new}(m)$
                \If {$w_{new}(m)$\, has\, elements\, outside\, the\, interval\, [-15,-4]}
                    \If {``negate" == 0}
                        \State $\mu \gets -\mu$
                        \State ``negate" $\gets 1$
                        \State Go back to step 9
                    \Else
                        \State $\mu \gets -\mu$
                        \State ``negate" $\gets 0$
                        \State Go back to step 8 for the next $\boldsymbol{W}(n)$
                    \EndIf
                \EndIf
                \State Calculate $P_{new}$ using updated $w_{new}(m)$
                \If {$P_{new} > P_{old}$}
                    \State $\boldsymbol{W} \gets \boldsymbol{W}_{new}$
                    \State $P_{old} \gets P_{new}$
                    \State $w(m) \gets w_{new}(m)$  for all $m$
                \Else
                    \State Perform the steps starting at line 12
                \EndIf
            \EndFor
            \State $\mu \gets \frac{\mu}{2}$
        \Until $\mu \geq 0.001$
    \end{algorithmic}
\end{algorithm}
\subsubsection{Simulation Results}
The Weight Ranking and Brute Force algorithms were implemented in MATLAB. The RIS configuration and the optimization goal are as follows. A spacing of 19 mm between the RIS elements is employed, $f_c=3$ GHz, and $f_b=\omega_b/(2\pi)= 12.9 $ MHz.
There are $M=100$ RIS elements and $N=50$ modes used to construct the voltage waveform. The array is extended by $2d_x$ at each of its ends, without placing varactors at these locations, simulating the waveform going through a longer path along the transmission line ($M_l = M_r = 2$). There is one desired receiver direction at $\theta^*=-30\degree$.

\begin{figure}[!t]
    \centering
    \includegraphics[width=1.0\linewidth]{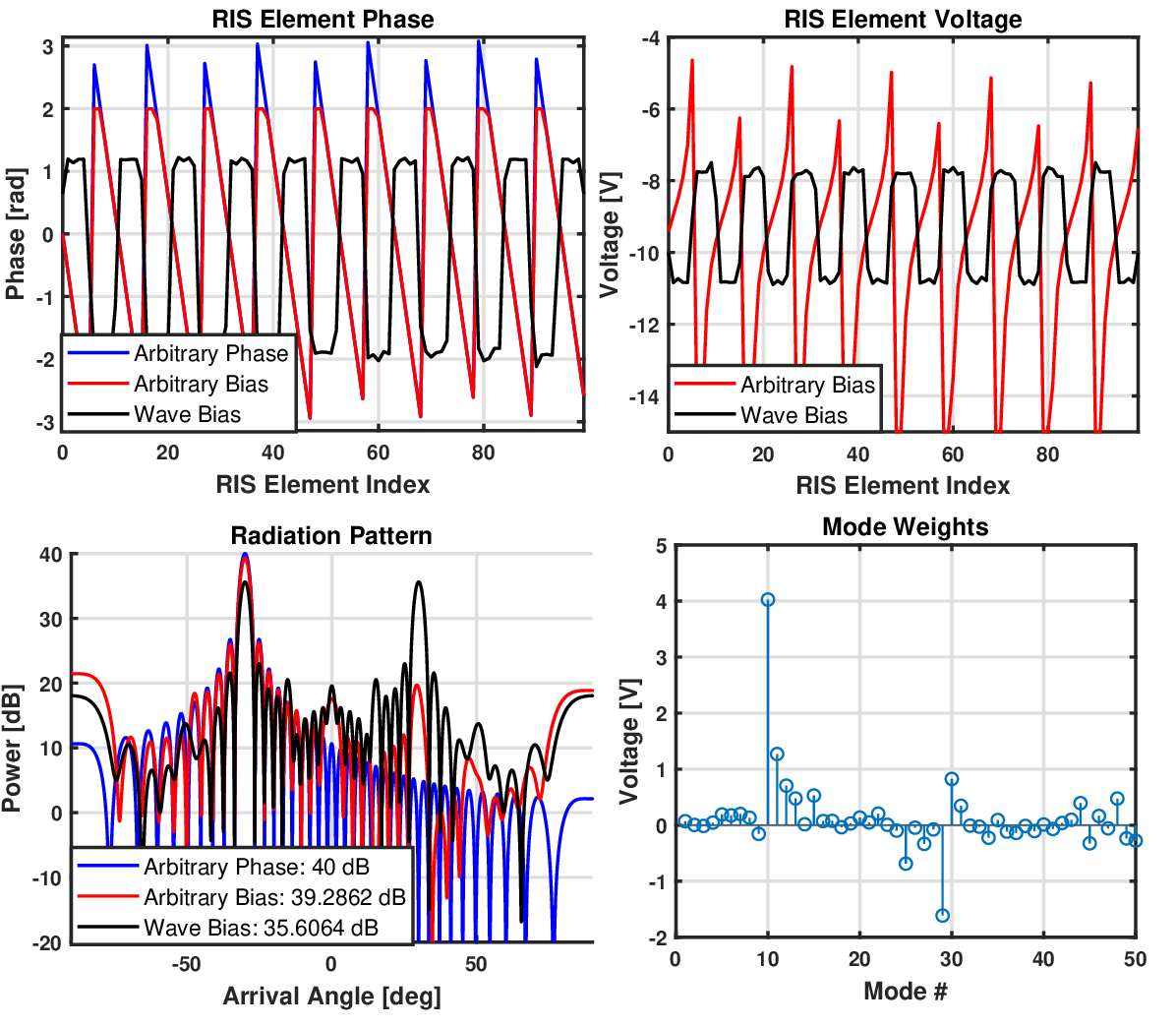}
    \caption{Maximizing SNR towards single receiver at $-30\degree$ using Weight Ranking and Brute Force algorithms.}
    \label{fig:ED_SNR_M30}
\end{figure}

One advantage of the Weight Ranking and Brute Force algorithms is that they do not require any prior knowledge on the shape of the desired voltage waveform to find an optimal set of weights, but rather they simply attempt to increase the SNR for a given desired direction. The results of this approach versus the ideal phase values and their corresponding arbitrary voltage values have been compared in Fig.~\ref{fig:ED_SNR_M30}.

\begin{figure}[!t]
    \centering
    \includegraphics[width=1\linewidth]{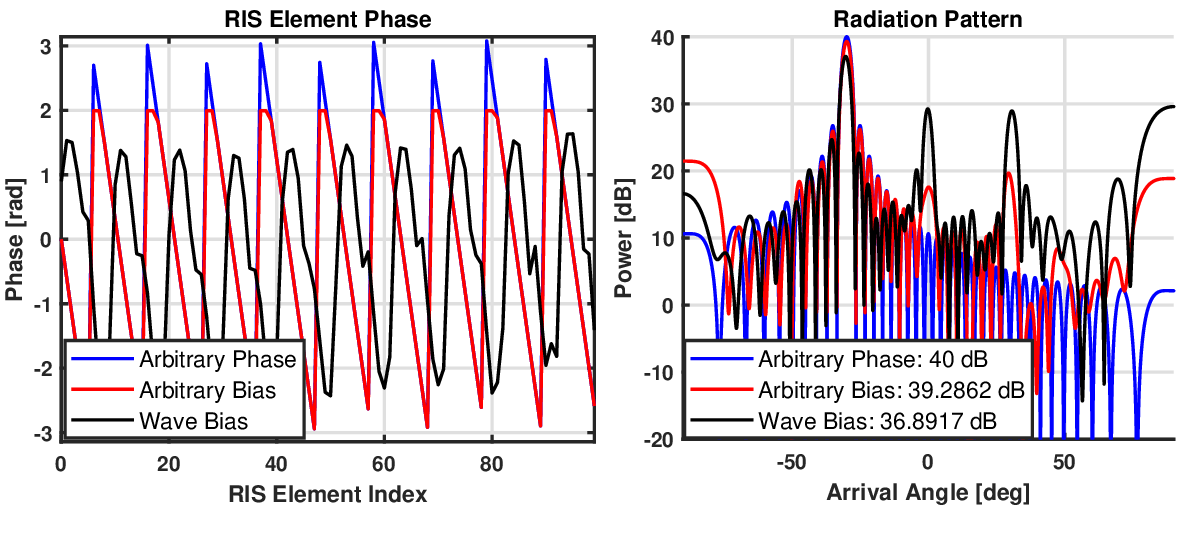}
    \caption{Maximizing SNR towards single receiver at $-30\degree$ using Weight Ranking and Brute Force algorithms, with $M_l = 2$ and $M_r = 102$.}
    \label{fig:ED_SNR_M30_104_DE}
\end{figure}

It can be seen in Fig.~\ref{fig:ED_SNR_M30} that there is a 3.7 dB loss between the model that uses the arbitrary voltage values and the wave-controlled approximation. The only resemblance between the
standing-wave model and the ideal models in the voltage and phase curves is that the spatial frequency of the standing waves matches;
otherwise the standing-wave model appears more like a square wave than the ideal sawtooth-shaped waveform. In addition to the desired peak at $-30^\circ$, there is a phantom peak at $30^\circ$ due to the symmetry
in the voltage waveforms, as the curves appear to be mirrored around the 50th RIS element. One way to eliminate this symmetry would be to double the length of the transmission line without adding more elements, but this would require doubling the length of the physical structure without increasing the SNR gain. When increasing the length by $2d_x$ on one side of the array and $102d_x$ on the other, the radiation pattern shown in Fig.~\ref{fig:ED_SNR_M30_104_DE} results, yielding a weaker peak at $30\degree$ that is approximately 8.4 dB lower than for the desired direction. However, the additional gain at $-30\degree$ compared to the previous case is only around 1.3 dB. Interestingly, the phase curve of the standing
wave model more closely resembles that of the ideal model, with some differences in the spatial phase shifts and amplitudes of the standing waves, likely due to the use of an insufficient number of high frequency components
to construct the waveform. A disadvantage of this algorithm is that not all of the modes are being fully employed; only a few of the modes have high-amplitude weights, while many others are near zero. This is an inherent weakness of the hill-climber algorithm as it tends to converge to a local minimum while not exploring different combinations of weights. Alternative algorithms such as Simulated Annealing may provide better performance.
\subsection{Solution for Optimization Problem 1: Sample-and-Hold Circuit}
\subsubsection{Standing Wave Bias - Using Sample-and-Hold Model}
%
A sample-and-hold (SH) circuit at each RIS element can be used to sample the standing wave voltage in~(\ref{eqn:wmt1}) at a specific time instant, and hold that voltage until the next sampling cycle. Assume that all RIS elements are sampled at the
same arbitrary time $t_0$ such that $\sin(n\omega_b t_0)\neq 0,$ $n= 1, 2, \ldots , N$. Then, the $\sin(n\omega_b t_0)$ terms in the modal expansion are no more than just weighting factors for each $W_n$, leading to
\begin{equation}
\begin{split}
    w & (m) = w(md_x,t_0)
    \label{eq:mmse_wave_model_dummy_elements}
\end{split}
\end{equation}

To match the wave modes with the arbitrary voltage waveform defined in 
Section~V-B.2, a Least Squares (LS) algorithm is derived below. As before, the length of the transmission line was extended at either end of the RIS to eliminate the boundary conditions and improve the match between the original waveform and that
generated by the limited modes.
For ease of implementation, the waveform generated by the sinusoids is centered around the average voltage of the arbitrary voltage bias $V(m)$ that would ideally be supplied to each varactor index $m$,
\begin{equation}
    W_0=\frac{1}{M}\sum_{m=0}^{M-1}{V(m)}.
    \label{eq:mmse_w0}
\end{equation}
Then, the variable $W_0$ is removed from the LS optimization and the objective function becomes
\begin{equation}
    \min_{\boldsymbol{W}}{J} = \min_{\boldsymbol{W}}\sum_{m=0}^{M-1}{||w(m)-V(m)||^2_2}
\end{equation}
where $\boldsymbol{W}=[W_1,W_2,\ldots,W_N]^T$ is the column vector containing the mode amplitudes. The LS solution is
\begin{equation}
\begin{split}
    \boldsymbol{W}=
    \left(\sum_{m=0}^{M-1}{\boldsymbol{s}_m^{} \boldsymbol{s}_m^T}\right)^{-1}
\left(\sum_{m=0}^{M-1}{(V(m)-W_0)\boldsymbol{s}_m}\right)
\label{eqn:WienerHopf1}
\end{split}
\end{equation}
where
\begin{equation}
\begin{split}
   \boldsymbol{s}_m = \ \ &\\
   \Big [ \sin  & \left(\frac{\pi (m+M_l)}{M-1+M_l+M_r}\right)\sin(\omega_b t_0),
\ldots , \\
& \sin\left(\frac{N\pi (m+M_l)}{M-1+M_l+M_r}\right)\sin(N \omega_b t_0) \Big ] ^T .
\end{split}
\end{equation}
Please refer to Appendix~A for the derivation of the algorithm.
\subsubsection{Sample-and-Hold Model Simulation Results}\label{sec:sample-and-hold-simulations}

\begin{figure}[!t]
    \centering
    \includegraphics[width=1.0\linewidth]{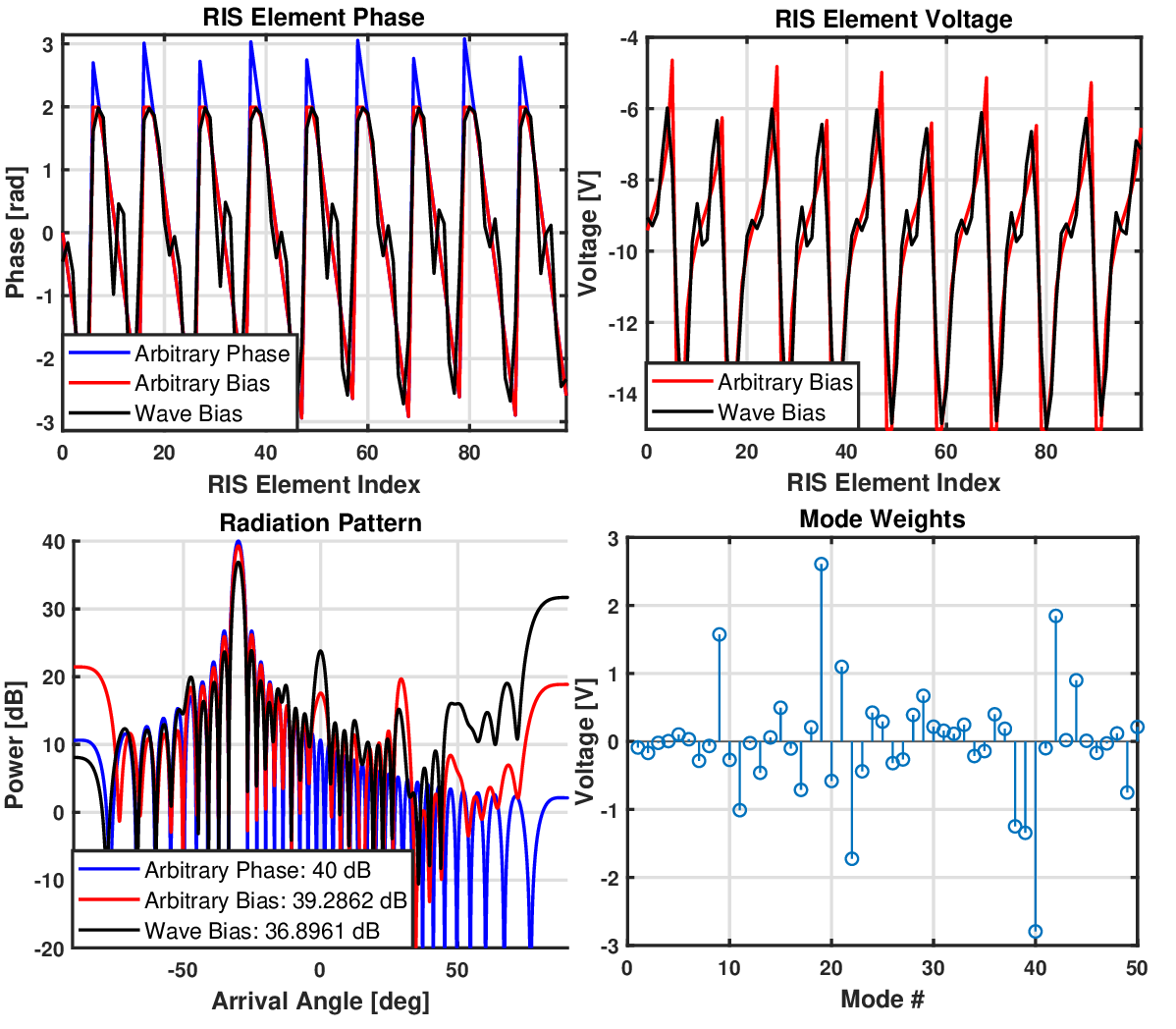}
    \caption{Maximizing SNR towards single receiver at $-30\degree$ with LS approach.}
    \label{fig:SH_SNR_M30_ORIG}
\end{figure}
We simulate the LS algorithm for an RIS with $M=100$ elements using $N=50$ sinusoidal modes. The variable $t_0$ was chosen as $\frac{8}{\omega_b}$ which guarantees that
$\sin(n\omega_b t_0)\neq 0,$ $n=1,2,\dots,50$. As before, we assume a far-field wave with normal incidence and a desired receiver located at an azimuth angle of $-30\degree$ from the RIS. To eliminate edge effects, the transmission line is $2d_x$ longer on either side of the board ($M_l=M_r=2$).

Fig.~\ref{fig:SH_SNR_M30_ORIG} plots the results for this case.
Compared to the radiation pattern generated using the arbitrary voltage bias, the wave-controlled approach has 2.4 dB less beamforming gain in the desired direction. The radiation patterns generally look similar and the spurious peaks at $0\degree$ and $30\degree$ are much smaller than in the case of the envelope detector. The performance of the LS algorithm can be improved by taking into account the fact that certain biasing voltages are more important for differentiating the RIS phase response. As seen in Fig.~\ref{fig:Phase_Ref_Coef}, especially for carrier frequencies of 2.9 and 3 GHz, the sensitivity of the phase is much higher for certain voltage ranges. For example, for 3 GHz, biasing voltages between -6V and -9V produce very large changes in the phase, while voltages less than -9V result in much less variation. As explained in the next section, this sensitivity can be exploited by weighting the importance of the biasing voltages in the LS optimization.
%

\subsubsection{Weighted LS}
\label{sec:UpdatedMMSE}
\begin{figure}[!t]
    \centering
    \includegraphics[width=0.7\linewidth]{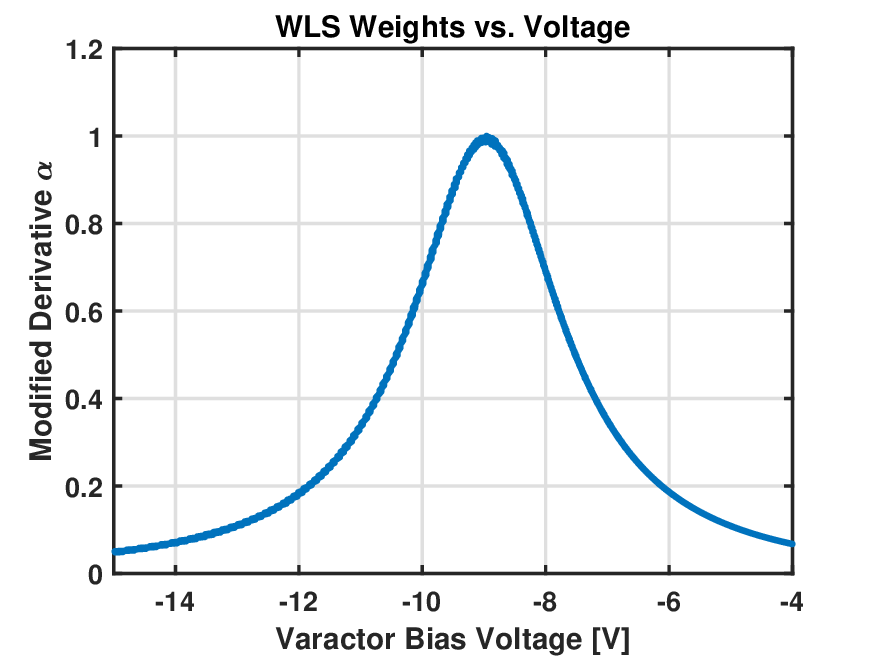}
    \caption{Modified phase versus voltage derivatives as weights for WLS.}
    \label{fig:phase-derivative}
\end{figure}

%
Clearly, the sensitivity of the phase to changes in the biasing voltage is reflected by the slope of the biasing curves in Fig.~\ref{fig:Phase_Ref_Coef}, and thus a reasonable way to assign the weights is based on this slope. First, we discretize the voltage values between -15V and -4V in 1mV steps and calculate the derivative of the reflection phases with respect to DC voltage bias. Then, we normalize the results between 0 and 1 and add 0.001 to each normalized derivative to eliminate possible zero weights. The final weight, $\alpha(m)$, corresponding to each RIS element $m$, is defined as
\begin{equation}
\begin{split}
    & \alpha(m)=\\
    & \frac{|\frac{\varphi(V(m))-\varphi(V(m)+0.001)}{0.001}|}{\max_{V\in\{-15,-14.999,...,-4\}}{|\frac{\varphi(V(m))-\varphi(V(m)+0.001)}{0.001}}|}+0.001,
    \label{eq:WLS_mod_deriv}
\end{split}
\end{equation}
Fig.~\ref{fig:phase-derivative} shows the weight for each discrete voltage value.

With the weighting $\alpha(m)$ designed above, the Weighted LS (WLS) becomes
    \begin{equation}
        \min_{\boldsymbol{W}}J = \min_{\boldsymbol{W}}{\sum_{m=0}^{M-1}{\alpha(m)||w(m)-V(m)||_2^2}},
    \end{equation}
with the solution
    \begin{equation}
        \boldsymbol{W}=\left(\sum_{m=0}^{M-1}\alpha(m) \boldsymbol{s}_m^{} \boldsymbol{s}_m^T\right)^{-1}
    \left(\sum_{m=0}^{M-1}\alpha(m) (V(m)-W_0)\boldsymbol{s}_m\right).
    \label{eqn:WLS}
    \end{equation}

\begin{algorithm}[!t]\caption{WLS Solution to Match Standing Wave Amplitudes with Voltage Curve}\label{alg:mmse}
    \begin{algorithmic}[1]
        \State Calculate $\phi(m)$ using (\ref{eq:ideal_phase}), $m=0,1,\ldots,M-1$
        \State Calculate $V(m)$ using (\ref{eq:phase_to_voltage}), $m=0,1,\ldots,M-1$
        \State Calculate weights $\alpha(m)$ for each of the $V(m)$ values using the modified derivatives from (\ref{eq:WLS_mod_deriv})
        \Repeat
            \State Calculate $\boldsymbol{W}$ using~(\ref{eqn:WLS})
            \State Calculate $w(m)$ using $\boldsymbol{W}$ and (\ref{eq:mmse_wave_model_dummy_elements})
            \If{$\min(w(m)) < -15V$}
                \State{$\alpha(m) \gets \alpha(m) \times 2$ at $m$ at $m$ where $w(m)=\min(w(m))$}
                \State{$V(m) \gets V(m) + 0.005$ at $m$ where $w(m)=\min{w(m)}$}
            \Else
                \If{$\max(w(m)) > -4V$}
                    \State{$\alpha(m) \gets \alpha(m) \times 2$ at $m$ where $w(m)=\max(w(m))$}
                    \State{$V(m) \gets V(m) - 0.005$ at $m$ where $w(m)=\max{w(m)}$}
                \EndIf
            \EndIf
        \Until{$w(m)$ has no elements outside the range [-15V, -4V]}
    \end{algorithmic}
\end{algorithm}

\begin{figure}[!t]
    \centering
    \includegraphics[width=1.0\linewidth]{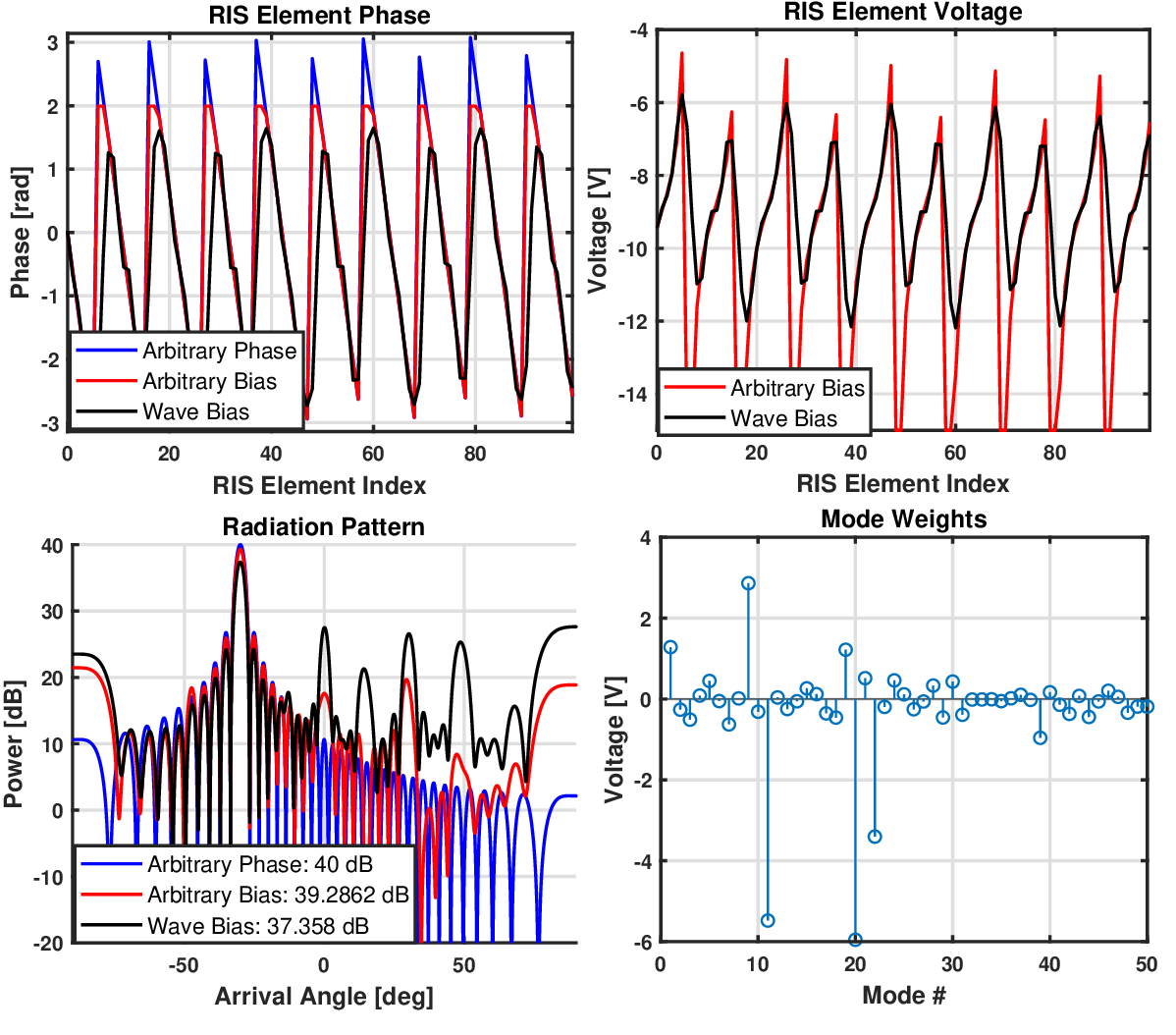}
    \caption{Maximizing SNR towards single receiver at $-30\degree$ with WLS approach.}
    \label{fig:SH_SNR_M30_MOD}
\end{figure}

The simulations were repeated for the same scenario as in 
Section~V-D.2 using the WLS approach, which is outlined in Algorithm~\ref{alg:mmse}. It is possible that the WLS solution will result in voltage values that exceed the -15V and -4V boundaries. Therefore, Algorithm~\ref{alg:mmse} provides additional steps that tighten the voltage boundaries every time this happens while increasing the weights associated with the element locations where the boundaries are violated. The $\boldsymbol{W}$ vector is calculated repeatedly until a solution is found that satisfies the original boundaries. The results shown in Fig.~\ref{fig:SH_SNR_M30_MOD} demonstrate much smoother voltage and phase curves and an improved beampattern with around 1.9 dB loss in beamforming gain compared with the ideal case.
%
\subsection{Comparisons of the Two Approaches for Optimization Problem~1}
The two varactor biasing configurations yield different performance for the case of maximizing SNR towards a single receiver direction. The main differences are in terms of the time required for convergence and the power reflected towards the desired direction. Simulation results for both approaches are compared in Table~\ref{tab:ed_vs_sh} for a case with $M=100$ RIS elements, $N=50$ modes and transmission lines extended $2d_x$ on the left and $2d_x$ on the right. The Weight Ranking and Brute Force algorithms were implemented for the envelope detector model, while the WLS algorithm was implemented for the sample-and-hold model. All simulations assume normal incidence to the RIS surface with  carrier frequency $f_c=3$ GHz and standing wave $f_b=12.9$ MHz.

\begin{table*}[!t]
    \centering
    \caption{Performance Comparisons Between SNR Maximization for Envelope Detector vs. Sample and Hold Models.}
    \begin{tabular}{||r||c|c||c|c||}
    \hline
        \multicolumn{1}{||c||}{Rx Dir.} & \multicolumn{2}{c||}{Envelope Detector} & \multicolumn{2}{c||}{Sample and Hold} \\
         $\theta^*$ & Power Steered & Optimization Time & Power Steered & Optimization Time \\
         \hline
         $-10\degree$ &  35.9418 dB & 209.261 s & 39.0365 dB & 0.361 s \\
         $-30\degree$ & 35.6064 dB & 198.845 s & 37.3580 dB & 0.369 s\\
         $-45\degree$ & 35.6419 dB & 180.316 s & 35.0566 dB & 0.380 s \\
         $-60\degree$ & 35.4927 dB & 183.326 s & 34.7838 dB & 0.365 s \\
         $-72\degree$ & 35.5045 dB & 204.159 s & 34.6151 dB & 0.348 s \\
         $24\degree$ & 35.4088 dB & 162.451 s & 37.9390 dB & 0.333 s  \\
    \hline
    \end{tabular}
    \label{tab:ed_vs_sh}
\end{table*}

The results demonstrate that the WLS optimization yields superior results compared to the Weight Ranking and Brute Force algorithms. Although all algorithms perform well in steering power
towards the desired receive, the WLS approach is faster by orders of
magnitude. Another weakness of the envelope detector model is its creation of a ``ghost'' peak in the negative of the desired direction, which can be avoided by making the transmission
line much longer and eliminating the symmetry of the standing waves where varactors are present, as seen previously in Fig.~\ref{fig:ED_SNR_M30}. On the contrary, the WLS approach does not share this artificial symmetry, as seen in Fig.~\ref{fig:SH_SNR_M30_MOD}.
Based on these results, we see that the sample-and-hold model can be optimized much more efficiently and create a more accurate radiation pattern. Moreover, as demonstrated in the next section, the WLS approach is very effective for the problem of optimizing for the SLNR when both beams and nulls must be created.
%
\subsection{Optimization Problem 2: Maximizing Signal-to-Leakage-plus-Noise Ratio at Multiple UE Directions}
As discussed previously, the SLNR problem is defined by (\ref{eq:SLNR_max}), repeated here
\begin{equation} \nonumber
    \max_{\boldsymbol{\eta}}{\slnr}=\max_{\boldsymbol{\eta}}{\frac{\min_{i\in\{1,2,\ldots,K\}}{\rho_s|\boldsymbol{h}_{d,i}^T\boldsymbol{\Phi g}|^2}}{\max_{j\in\{{1,2,\ldots,L\}}}{\rho_s|\boldsymbol{h}_{e,j}^T\boldsymbol{\Phi g}|^2}+\sigma_s^2}}.
\end{equation}
The goal is to maximize the power reflected towards all $K$ intended receiver directions, while minimizing the power reflected towards all $L$ undesired receiver directions (either minimizing eavesdropping or reducing unwanted interference). The optimization metric is defined by the ratio of the minimum power directed towards a desired receiver and the maximum power steered towards an undesired direction with additive noise.

For this task, the Weight Ranking and Brute Force algorithms are not directly applicable, due to the difficultly in computing gradients for~(\ref{eq:SLNR_max}).
Instead, we develop an alternative algorithm based on
Simulated Annealing (SA) that will be discussed further in 
Section~V-F.2. Afterwards, simpler analytical algorithms that can be employed using the sample-and-hold circuit model based on WLS optimization will be discussed as well.
Before discussing the SA approach, it is necessary to discuss an important feature about the relationship between the standing waves and the corresponding radiation pattern of the RIS. Since SA requires random searches from a specific starting point, it is crucial to determine the best initialization for faster and more accurate convergence, similar to how the Weight Ranking algorithm provides an initial order for tuning the modes one-by-one. However, the approach presented below for SA is more intuitive and analytical, and results in a much simpler method for initializing the weights for further optimization.
\subsubsection{Correlation Between Modal Frequencies and Peaks in the Radiation Pattern}
As discussed above, further investigation of the relationship between the individual modes and the radiation pattern generated by the RIS is required to derive a more efficient optimization algorithm. Referring back to Fig.~\ref{fig:RIS_Single_Mode}, it was demonstrated that a single sinusoid can produce two peaks at $\pm\theta^*$ in the radiation pattern for the envelope detector model. This is the result of the phase shift gradient across the RIS that collectively reflects a beam towards a specific direction \cite{10354402}. If the phase gradient is steeper, corresponding to a sinusoid with higher frequency, the absolute value of the reflection angle increases. The same effect was seen in the sample-and-hold model. The expression for the mode number $n$ that generates peaks at $\pm\theta^*$ for the sample-and-hold model is given by
\begin{equation}
    n_{S/H}=\round{|2(M+1)\Delta\sin(\theta^*)|},
    \label{eq:main_mode_index}
\end{equation}
where $M$ is the number of RIS elements, $\Delta$ is the distance in wavelengths between the RIS elements, and $n$ is rounded to the nearest integer value via the function $\lfloor\, .\, \rceil$ since the standing-wave modes are discrete. This formula also suggests that the spatial frequency corresponding to $n$ is the minimal mode frequency required to generate a peak at $\theta^*$. The derivation is provided in Appendix B.

A slightly different model holds for the envelope detector model. The transmission line assumed in our model in \cite{main-ris} is terminated by a short circuit. Therefore, the voltage reflection coefficient at the end of the transmission line is $\Gamma=-1$, and the reflected wave is inverted at the boundary \cite{smith-chart}. Thus, each point on the transmission line experiences peaks due to both the positive and the negative traveling waves, and thus samples twice the number of peaks since it samples absolute values. Therefore, for a standing wave oscillating at frequency $f$, the peak detector will sample a peak at frequency $2f$, and the expression for the mode index $n_{PD}$ that corresponds to the peak at $\theta^*$ will be
\begin{equation}
    n_{PD}=\frac{n_{S/H}}{2} .
    \label{eq:main_modes_envelope_detector}
\end{equation}
This provides intuition for initializing which weights should be optimized in the SA approach described next.

\subsubsection{Mode Amplitude Optimization Using Simulated Annealing}\label{sec:ModeAmplitudeOptimization}
When optimizing a non-convex objective over a large number of variables, many algorithms tend to settle on local minima that may be far from the global optimum \cite{nonlinear-observations}. Particularly when using hill-climber methods such as the previously described combination of Weight Ranking and Brute Force search, the solution for the weights is highly dependent on the initialization, as well as the order in which the weights are being solved for. To address this issue, we use the  Simulated Annealing approach described below.

Simulated Annealing (SA) is a stochastic optimization method that employs randomization to increase the likelihood of convergence to the global optimum. SA relies on the principle of ``annealing'' from physics, where a solid is cooled until it reaches its minimal energy state \cite{traveling-salesman-thermo}. SA uses Boltzmann distributions to find the probability of a state based on its temperature $T>0$ and energy $f(x)$. The algorithm works as follows: Start with an initial system state and temperature. As the system matures, iteratively experiencing random updates that bring it towards a better or worse state with some probability that depends on the energy and temperature, the temperature decreases and approaches zero. As this happens, the system becomes less likely to randomly jump to worse states and converges towards a nearby minimum by moving in the direction that decreases its energy, which serves as the cost function \cite{simulated-annealing-ris}.

To implement SA for the SLNR optimization problem, assume a set of $K$ angles towards which the power gain should be maximized: ${\theta}^*_{arr} = [\theta^*_{d,1}, \theta^*_{d,2}, \ldots, \theta^*_{d,K}]$, and $L$ angles towards which the reflected beam should be minimized. Define the vector $\boldsymbol{W}$ representing the mode amplitudes as the ``state" of the system, and $\slnr$ and $\slnr_{new}$ as the ``energy" of the system before and after a state update, respectively. Define the probability of switching to the next state as
\begin{equation}
    p=\begin{cases}
         1 & \slnr_{new} > \slnr \\
         e^{\left(-\frac{\slnr-\slnr_{new}}{k_c T}\right)} & \slnr_{new} \leq \slnr.
    \end{cases}
    \label{eq:simulated_annealing_prob}
\end{equation}
Let $T$ denote the ``temperature" based on the current iteration of the algorithm and $k_c$ a constant representing the ``cooling factor." The initial state vector $\boldsymbol{W}$ is excited only at specific modes corresponding to the peak directions determined by~(\ref{eq:main_mode_index}), with amplitudes set to $3/K$. This initializes the algorithm to a good starting point, while allowing for enough margin to update all the mode amplitudes as the algorithm progresses without saturating the voltage limits. The next state $\boldsymbol{W}_{new}$ is determined by adding a Gaussian random variable $\epsilon$ drawn from  $\mathcal{N}(0,1)$ and scaled by a factor $\lambda$ to each of the amplitudes in the state vector $\boldsymbol{W}$. The updated DC voltages $w(m)$ at each RIS element are calculated. If the $\slnr_{new}$ of the new state is better than the current $\slnr$, then the algorithm chooses the better amplitude state. If it is worse, the algorithm will only update to that state if a random sample from a uniform distribution on the interval (0,1) is less than $p$, otherwise it will remain in the current state. Additionally, if the system remains in a worse state for longer than some upper limit of iterations, it will return to its previous best state $\boldsymbol{W}_{best}$ corresponding to $\slnr_{best}$ and continue from there. The details of our SA implementation are given in Algorithm~4.

\begin{figure}[!t]
    \centering
    \begin{subfigure}{0.49\textwidth}
    \centering
        \includegraphics[width=1.0\textwidth]{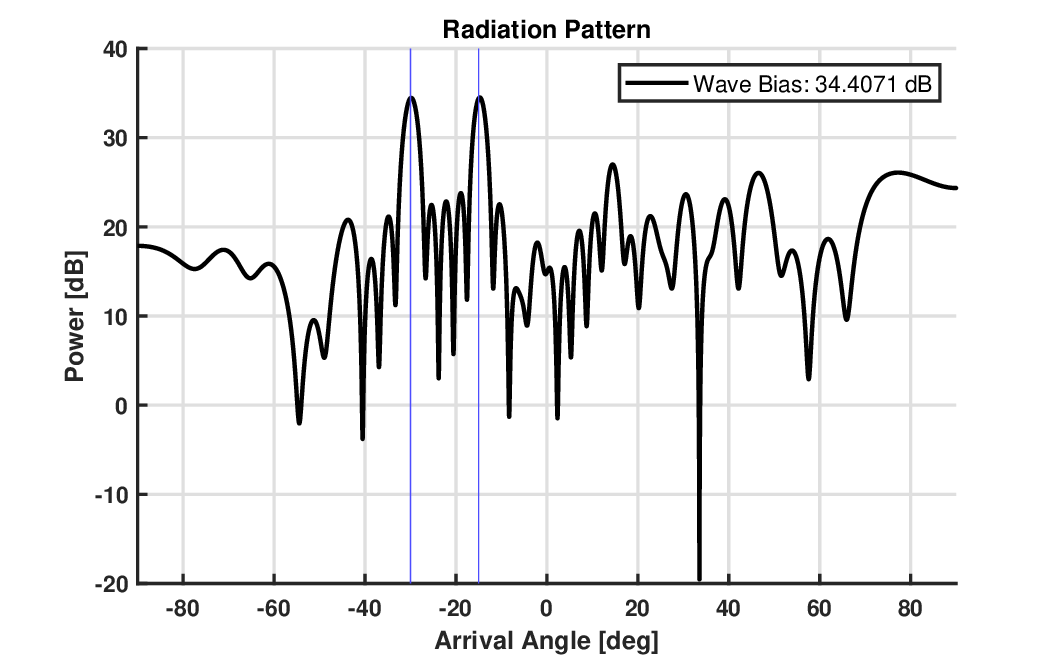}
        \caption{Radiation Pattern, SLNR = 34.41 dB}
    \end{subfigure}
    \begin{subfigure}{0.49\textwidth}
    \centering
        \includegraphics[width=1.0\textwidth]{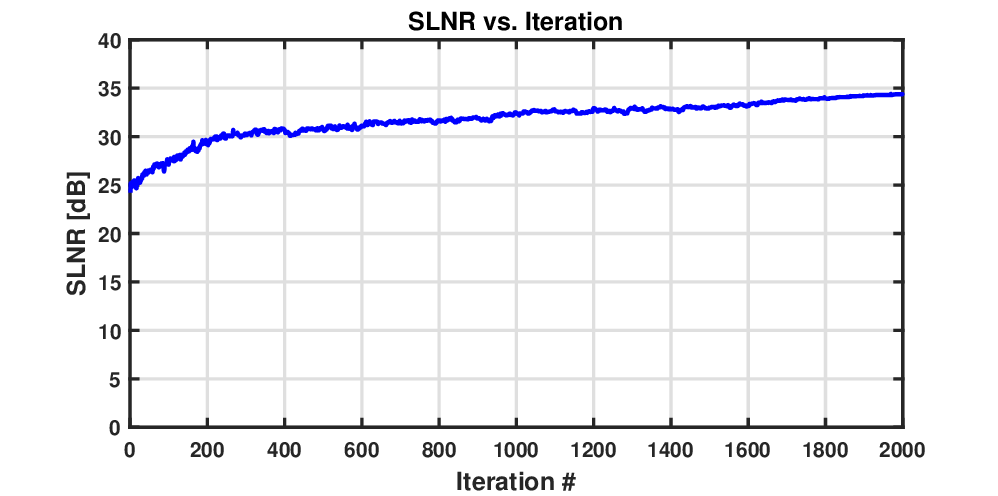}
        \caption{SLNR at every iteration of SA}
    \end{subfigure}
    \caption{Simulation results using SA for the sample-and-hold model, with desired receivers at $-30\degree$ and $-15\degree$ and no undesired receivers.}
    \label{fig:SH_SLNR_M15_M30_SA}
\end{figure}

\begin{figure}[!t]
    \centering
    \begin{subfigure}{0.49\textwidth}
    \centering
        \includegraphics[width=1.0\textwidth]{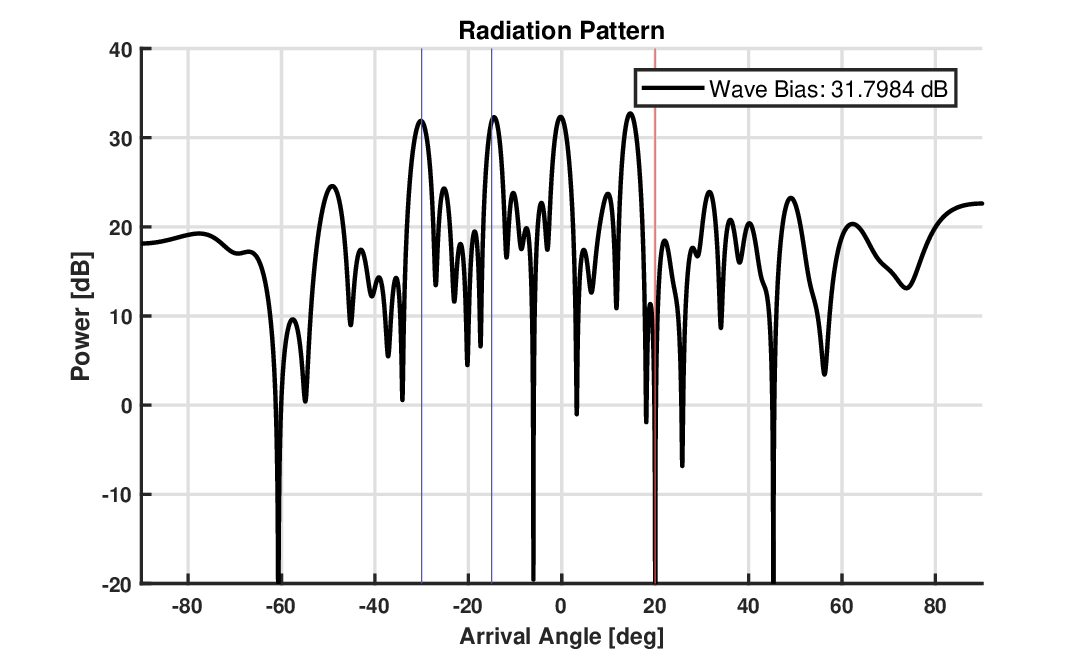}
        \caption{Radiation Pattern, SLNR = 31.80 dB}
    \end{subfigure}
    \begin{subfigure}{0.49\textwidth}
    \centering
        \includegraphics[width=1.0\textwidth]{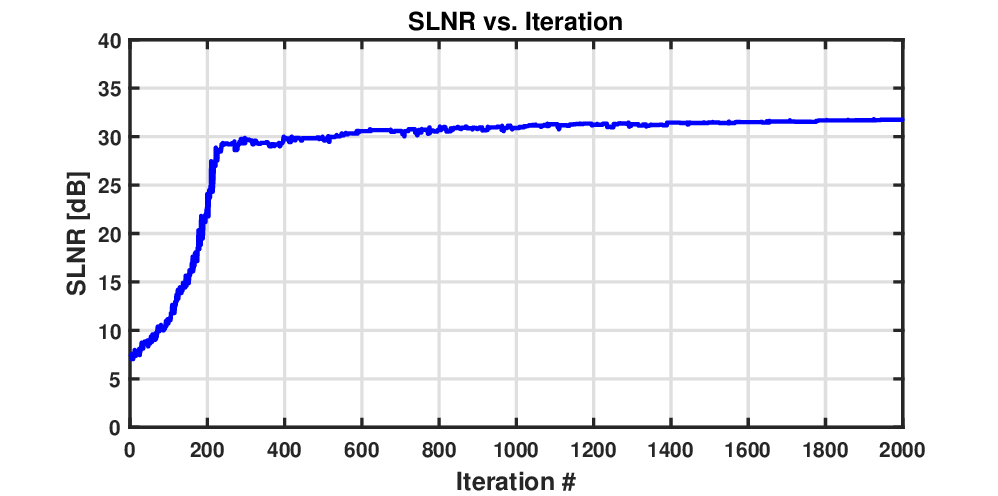}
        \caption{SLNR at every iteration of SA}
    \end{subfigure}
    \caption{Simulation results using SA for the sample-and-hold model, with desired receivers at $-30\degree$ and $-15\degree$ and undesired receiver at $20\degree$.}
    \label{fig:SH_SLNR_M15_M30_P20_SA}
\end{figure}

\begin{algorithm}[!t]\caption{Simulated Annealing}\label{alg:simulated_annealing}
    \begin{algorithmic}[1]
        \State $\boldsymbol{W} \gets [0,0,\ldots,0]^T$
        \State Calculate indices $n_k$ of each peak in $\theta^*_{arr}$, using (\ref{eq:main_mode_index})
        \State Each $\boldsymbol{W}(n_k) \gets \frac{3}{K}$
        \State $\slnr_{best} \gets -\infty$
        \State $\boldsymbol{W}_{best} \gets \boldsymbol{W}$
        \State $i_{best} \gets 0$
        \State Calculate initial $w(m)$ using $\boldsymbol{W}$ and~(\ref{eq:mmse_wave_model_dummy_elements}), $m=0,...,M-1$
        \State Calculate $\slnr$ [dB] using $w(m)$ and~(\ref{eq:SLNR_max})
        \For {$i\in\{1,2,\ldots,i_{max}\}$}
            \If {$i-i_{best}\geq 100$}
                \State $\boldsymbol{W} \gets \boldsymbol{W}_{best}$
                \State $i_{best} \gets i$
                \State $\slnr \gets \slnr_{best}$
            \EndIf
            \State $T \gets 100\left(1-\frac{i}{i_{max}}\right)$
            \State $\boldsymbol{W}_{new}(n)\! \gets \! \boldsymbol{W}(n) \! + \! \lambda \epsilon,\; \epsilon\! \sim\! \mathcal{N}(0,1), \ n= 1,2,\ldots,N$
            \State Calculate $w(m)$ using $\boldsymbol{W}_{new}$ and~(\ref{eq:mmse_wave_model_dummy_elements}), $m=0,...,M-1$
            \If {$w(m)$ has elements outside [-15V, -4V]}
                \State Increment $i$, go to step 10
            \EndIf
            \State Calculate $\slnr_{new}$ using $w(m)$ and~(\ref{eq:SLNR_max})
            \If{$\slnr_{new} > \slnr_{best}$}
                \State $\slnr_{best} \gets \slnr_{new}$
                \State $\boldsymbol{W}_{best} \gets \boldsymbol{W}$
                \State $i_{best} \gets i$
            \Else
                \State Calculate $p$ using~(\ref{eq:simulated_annealing_prob})
                \If{$p \geq$ rand(1)}
                    \State $\boldsymbol{W} \gets \boldsymbol{W}_{new}$
                    \State $\slnr \gets \slnr_{new}$
                \EndIf
            \EndIf
        \EndFor
    \end{algorithmic}
\end{algorithm}
In our simulations, the SA algorithm was implemented with $\lambda=0.03$, cooling factor $k_c=0.002$, maximum number of iterations $i_{max}=2000$. For the example in Fig.~\ref{fig:SH_SLNR_M15_M30_SA}, two beams at $-30\degree$ and $-15\degree$ are desired assuming $M=100$ RIS elements and $N=50$ modes, with the transmission line extended by $2d_x$ before and after the first and last  varactor ($M_l = M_r = 2$). The sample-and-hold circuit model was used for this example. The SA algorithm was able to increase the gain by almost 10 dB from the initialization point and achieves strong beams in the desired receiver directions, without reflections towards $15\degree$ and $30\degree$. Fig.~\ref{fig:SH_SLNR_M15_M30_P20_SA} shows the results for the same case as in Fig.~\ref{fig:SH_SLNR_M15_M30_SA}, except that a desired null is added at $20\degree$. The SA algorithm improves the SLNR by around 25 dB and provides a deep null towards $20\degree$, albeit at the cost of higher sidelobes in other directions.

\begin{figure}[!t]
    \centering
    \begin{subfigure}{0.49\textwidth}
    \centering
        \includegraphics[width=1.0\textwidth]{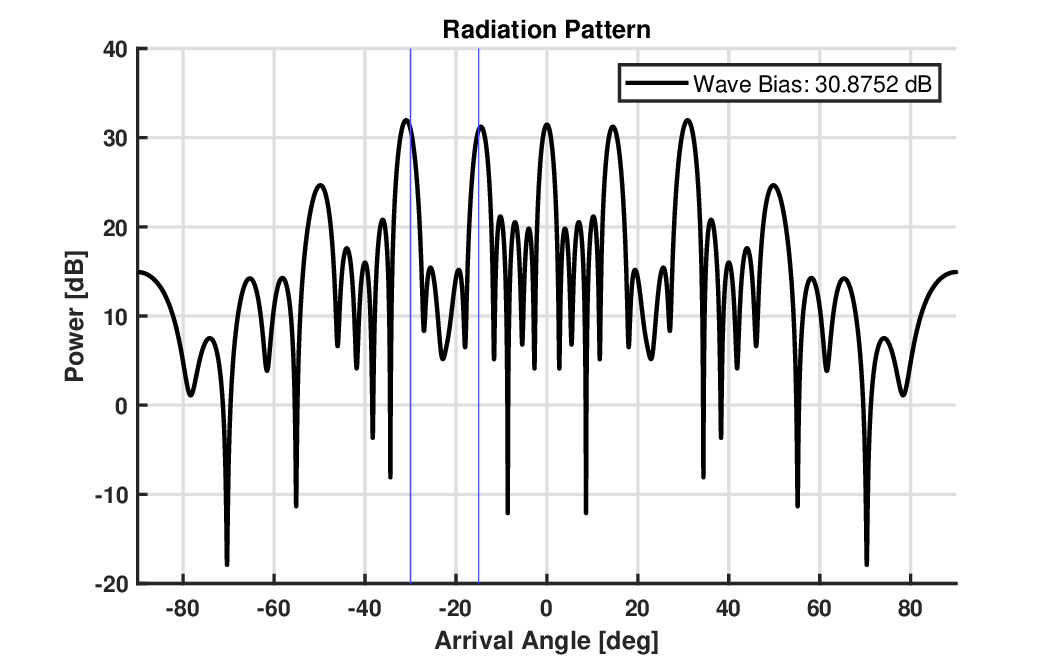}
        \caption{Radiation Pattern, SLNR = 30.88 dB}
    \end{subfigure}
    \begin{subfigure}{0.49\textwidth}
    \centering
        \includegraphics[width=1.0\textwidth]{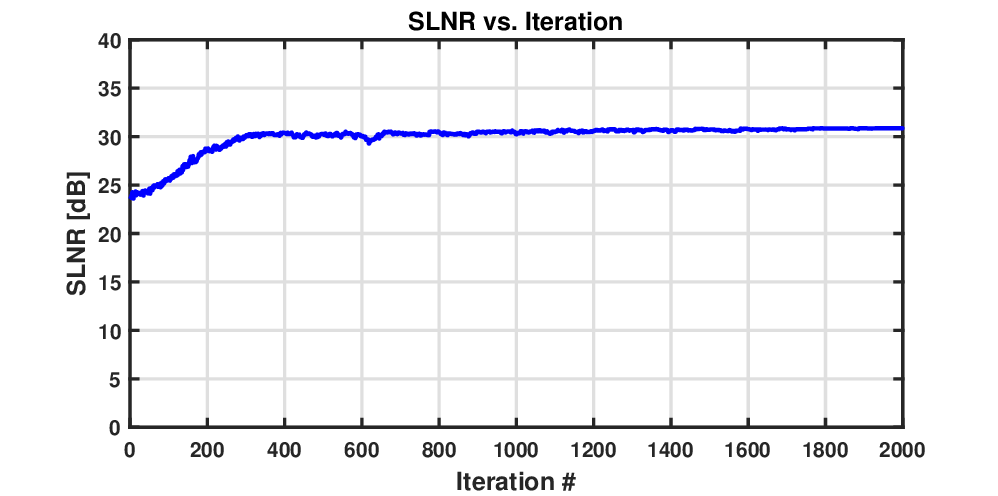}
        \caption{SLNR at every iteration of SA}
    \end{subfigure}
    \caption{Simulation results using SA for the envelope detector model, with desired receivers at $-30\degree$ and $-15\degree$ and no undesired receivers.}
    \label{fig:ED_SLNR_M15_M30_SA}
\end{figure}

\begin{figure}[!t]
    \centering
    \begin{subfigure}{0.49\textwidth}
    \centering
        \includegraphics[width=1.0\textwidth]{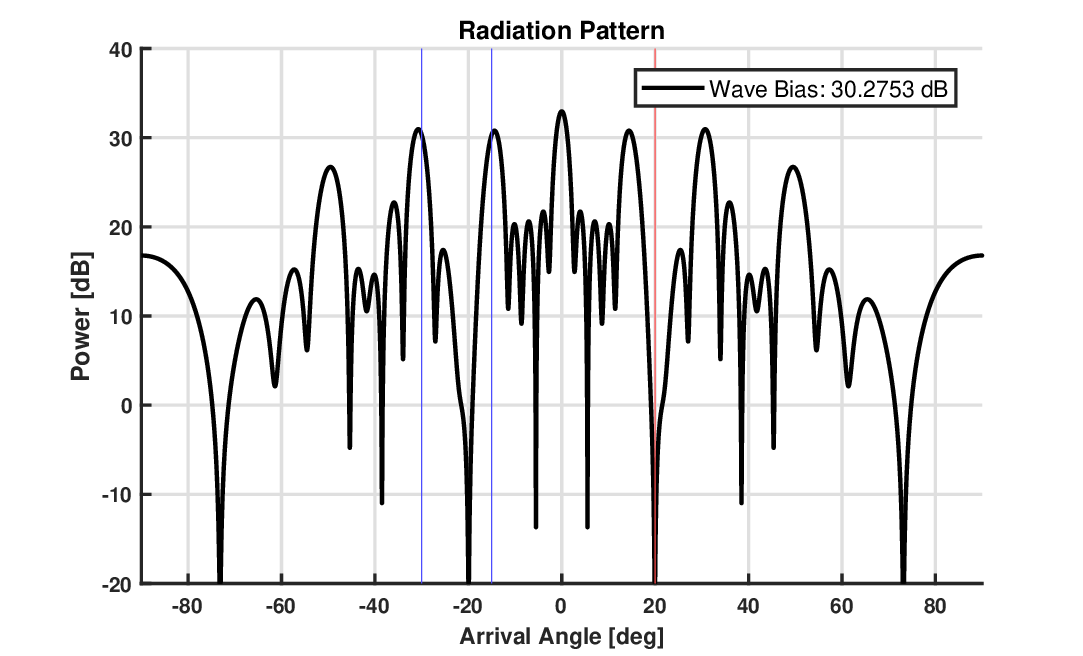}
        \caption{Radiation Pattern, SLNR = 30.28 dB}
    \end{subfigure}
    \begin{subfigure}{0.49\textwidth}
    \centering
        \includegraphics[width=1.0\textwidth]{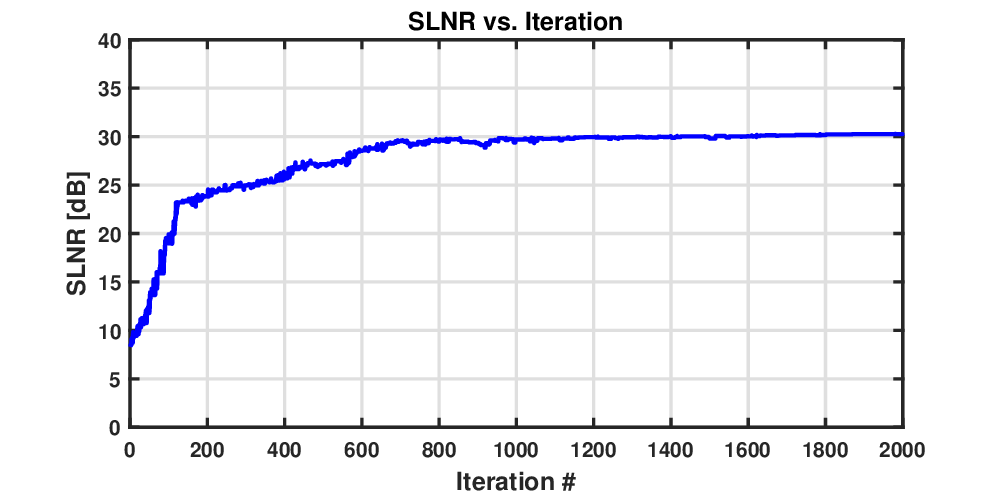}
        \caption{SLNR at every iteration of SA}
    \end{subfigure}
    \caption{Simulation results using SA for the envelope detector model, with desired receivers at $-30\degree$ and $-15\degree$ and an undesired receiver at $20\degree$.}
    \label{fig:ED_SLNR_M15_M30_P20_SA}
\end{figure}

We next applied the SA algorithm for the case of the envelope detector model with $W_0=-4$V and $w(m)$ calculated according to (\ref{eq:envelope_detector_wave_model}). The same simulation parameters were used as in the previous case, except that initial mode indices $n_k$ were calculated instead using~(\ref{eq:main_modes_envelope_detector}). The simulation results are shown in Fig.~\ref{fig:ED_SLNR_M15_M30_SA} and Fig.~\ref{fig:ED_SLNR_M15_M30_P20_SA}. We see that in both cases, the SA algorithm provides a significant boost in SLNR of approximately 6 dB and 20 dB, and is able to form deep nulls in directions close to the main beams. As in previous examples, the envelope detector architecture produces higher sidelobes and a strong beam in the broadside direction, unlike the sample-and-hold approach.

In the next section we focus on heuristic approaches for designing the RIS response to simultaneously steer beams and nulls in certain directions.
%
\subsubsection{Maximizing Power Towards Multiple Receiver Directions Analytically}
We begin with the problem of maximizing the power steered towards multiple receiver directions, without any nulls:
\begin{equation}
    \max_{\boldsymbol{\eta}}\min_{i\in\{1,2,\ldots,K\}}{\rho_s|\boldsymbol{h}_{d,i}^T\boldsymbol{\Phi g}|^2}.
\end{equation}
While even this simpler problem cannot be solved analytically for our two circuit models, an approximate solution can be found in a straightforward way, as discussed below for the three different parameterizations for $\boldsymbol{\eta}$.

\textit{Ideal Phase\/} -- Here we use~(\ref{eq:ideal_phase}) to find the optimal set of reflection coefficients for each individual receiver direction $\theta^*_{d,i}$ for $i=1,2,\cdots,K$. We refer to each of these RIS phase configurations as ${\phi_{d,i}}(m)\text{ for } m=0,1,\ldots,M-1$. Then, for each $m$, we take the average value of the reflection coefficients (which are complex), and we find the average over the $K$ solutions: 
\begin{equation}
        \phi(m)=  \frac{1}{K}\sum_{i=1}^{K}{{\phi_{d,i}}(m)} \\
\label{eqn:phi_m}
\end{equation}
To satisfy the unit amplitude constraint after the averaging, we simply keep just the phase of the result: $\phi(m) \leftarrow \exp{(j\phase{\phi(m)})}$.

\textit{Arbitrary Voltage Bias\/} -- 
As in 
Section~V-B.2, we take the ideal reflection coefficients calculated above and map them to voltage values using (\ref{eq:phase_to_voltage}), ensuring that the phase values remain within the boundaries allowed by the varactor biasing voltage.

\textit{Wave-Controlled Bias\/} -- 
We calculate the voltages of the modal decomposition using the WLS algorithm in 
Section~V-D.3.

\begin{figure}[!t]
    \centering
    \includegraphics[width=1\linewidth]{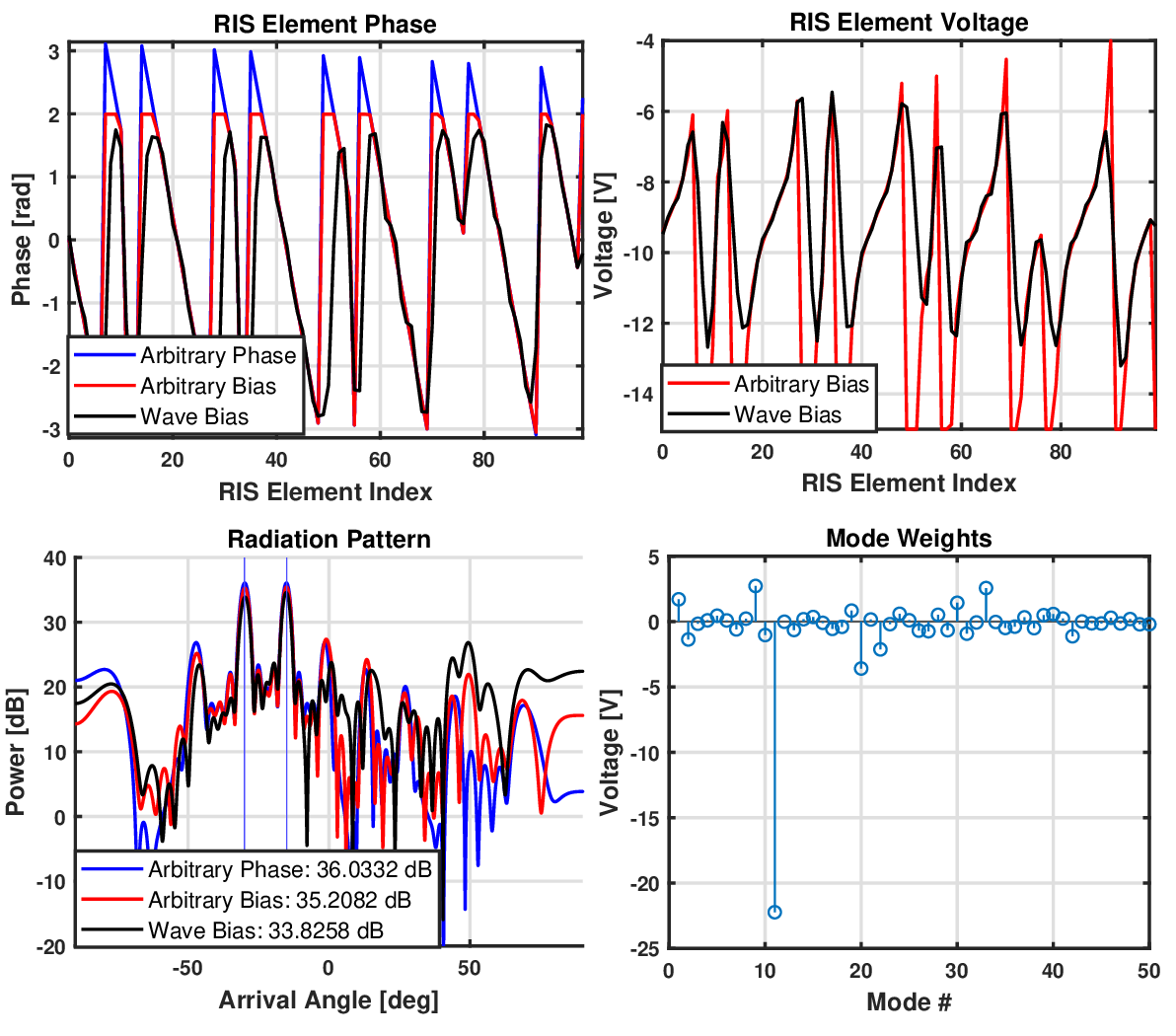}
    \caption{Maximizing power reflected towards two directions using the sample-and-hold model. $M=100$ RIS elements, $N=50$ modes, $M_l=M_r=2$. Desired beams at $-30\degree$ and $-15\degree$.}
    \label{fig:SH_SLNR_M30_M15_MMSE}
\end{figure}

\begin{figure}[!t]
    \centering
    \includegraphics[width=1\linewidth]{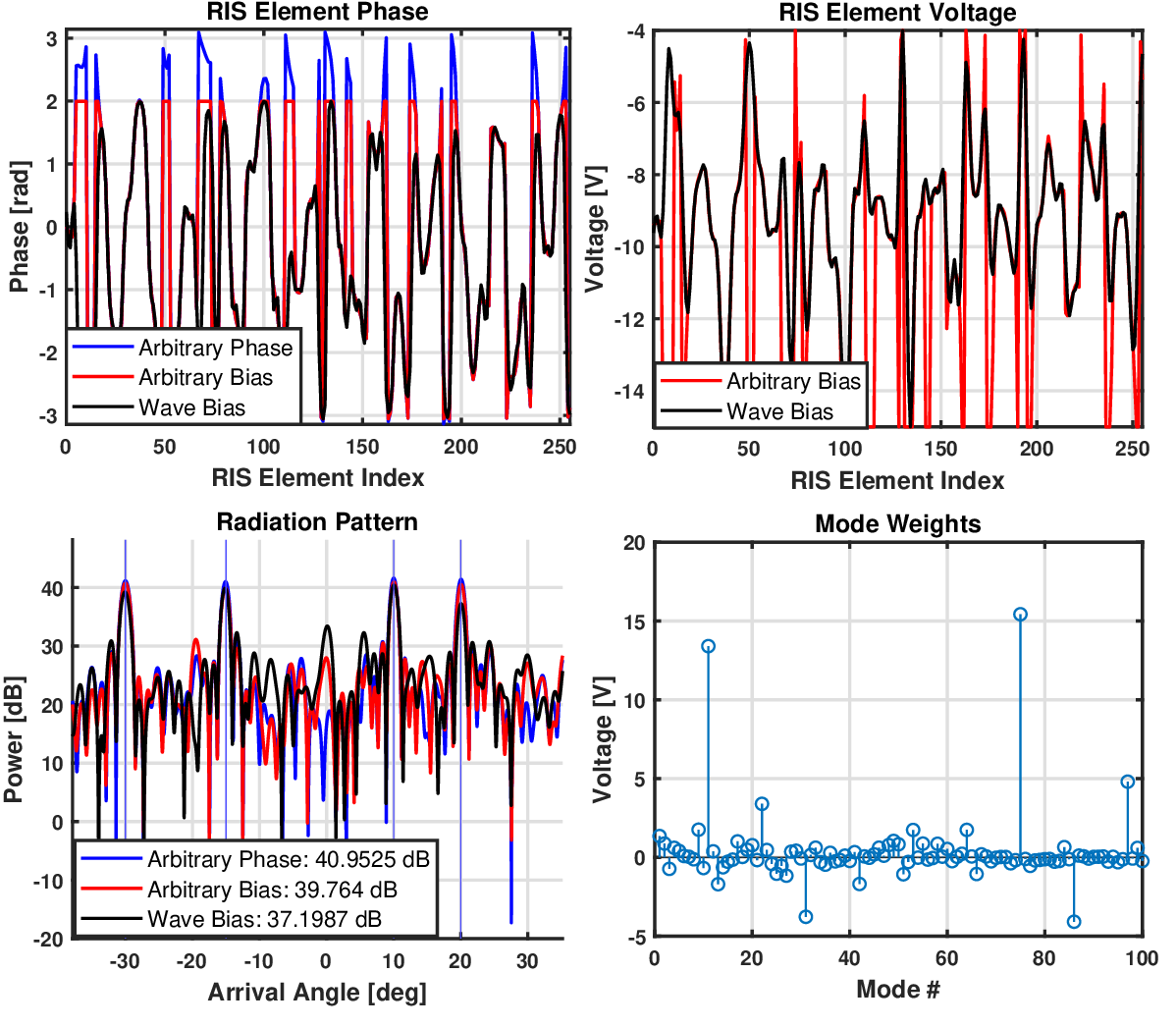}
    \caption{Maximizing power reflected towards four directions using sample-and-hold model. $M=256$ RIS elements, $N=100$ modes, $M_l=M_r=2$. Desired beams at $-30\degree, -15\degree, 10\degree,$ and $20\degree$.}
    \label{fig:SH_SLNR_M30_M15_10_20_MMSE}
\end{figure}

Taking the average of the reflection coefficients will in general ensure that all receivers receive approximately the same amount of power. Some simulation results confirming the effectiveness of the above approach are shown in Fig.~\ref{fig:SH_SLNR_M30_M15_MMSE} and~\ref{fig:SH_SLNR_M30_M15_10_20_MMSE}. We see that the simple averaging approach provides beams in the desired directions, while using the arbitrary voltage bias reduces the power by only about 1 dB, and the standing wave bias by another 1-3 dB. The next section considers the problem of simultaneous null- and beamsteering.

%
%
\subsubsection{Simultaneous Beam- and Null-Steering}
To form a null in a given direction $\theta^*_{e,j}$, the RIS configuration should satisfy
\begin{equation}
\label{eq:null}
    \sum_{m=0}^{M-1}\phi(m) e^{-jm\kappa(\theta^*_{e,j})} \approx 0.
\end{equation}
We propose a heuristic iterative approach that takes the solution from the previous section for the desired beams, and modifies it to add the nulls. The required steps are outlined in Algorithm~\ref{alg:slnr_ideal_phase} for the ideal phase case. The algorithm starts by calculating the reflection coefficients required to form beams at the desired directions. Then, it iteratively tunes the reflection coefficients by calculating their product with the channel coefficients that correspond to each null direction $\theta^*_{e,j}$,
\begin{equation}
    r_{j}(m)=\phi(m) e^{-jm\kappa(\theta^*_{e,j})},
\end{equation}
calculating the average value $\Bar{r}_j$, and subtracting the average from each $r_{j}(m)$ to make their new average zero. At this point,~(\ref{eq:null}) is satisfied and the updated reflection coefficients are mapped back into $\phi(m)$ by dividing the result by the channel coefficients and using the phase of the new result, as outlined in steps 10 and 11 of Algorithm~\ref{alg:slnr_ideal_phase}. The same procedure is repeated until the response of the RIS is orthogonalized towards all eavesdropper directions and the power gains at those directions are below some threshold $\mu$.


\begin{algorithm}[!t]
    \begin{algorithmic}[1]
        \For{each desired beam direction $\theta_{d,i}^*$}
            \State Calculate ${\phi_{d,i}}(m)$ using (\ref{eq:ideal_phase}), $\ i=1,\ldots,K$.
        \EndFor
        \State $\phi(m) \gets \frac{1}{K}\sum_{i=1}^{K} \phi_{d,i}(m)$.
        \State $\phi(m) \leftarrow \exp{(j\phase{\phi(m)})}$.
        \Repeat
            \For{each null direction $\theta_{e,j}^*$}
                \State $r_{j}(m) \gets \phi(m) e^{-jm\kappa(\theta_{e,j}^*)},\ m=0,\ldots,M-1$.
                \State Calculate $\Bar{r}_j=\frac{1}{M}\sum_{m=0}^{M-1} r_{j}(m)$.
                \State $\phi(m) \gets
               \left(\frac{r_{j}(m)-\Bar{r}_j}{e^{-jm\kappa\left(\theta_{e,j}^*\right)}}\right),\ m=0,\ldots,M-1$.
                \State $\phi(m) \gets \exp\left(j\phase{\phi(m)}\right),\ m=0,\ldots,M-1$.
            \EndFor
        \Until{$\max(|\Bar{r}_1|,\Bar{r}_2,\ldots,|\Bar{r}_L|) \leq \mu$}.
    \end{algorithmic}\
    \caption{RIS Design for Simultaneous Beam and Null Steering}
    \label{alg:slnr_ideal_phase}
\end{algorithm}

The same approach can be used for the arbitrary voltage case, except that the reflection phase values are converted to voltages and vice versa between iterations to account for the limited phase values. Since the changes in the phase and voltage curves are so slight, the WLS algorithm has trouble forming the nulls for the wave-controlled case and thus fails to match the SLNR values simply by attempting to match the ideal voltage and phase curves with the wave-controlled ones. For the final touches, we define the \textit{Combined Algorithm} -- Start with Algorithm~\ref{alg:slnr_ideal_phase} to find the initial reflection phases. Then, map those phase values into voltage values using~(\ref{eq:phase_to_voltage}) and convert those to mode amplitudes using WLS. Finally, increase the SLNR and form deep nulls using SA. 

\begin{figure}[!t]
    \centering
    \includegraphics[width=1\linewidth]{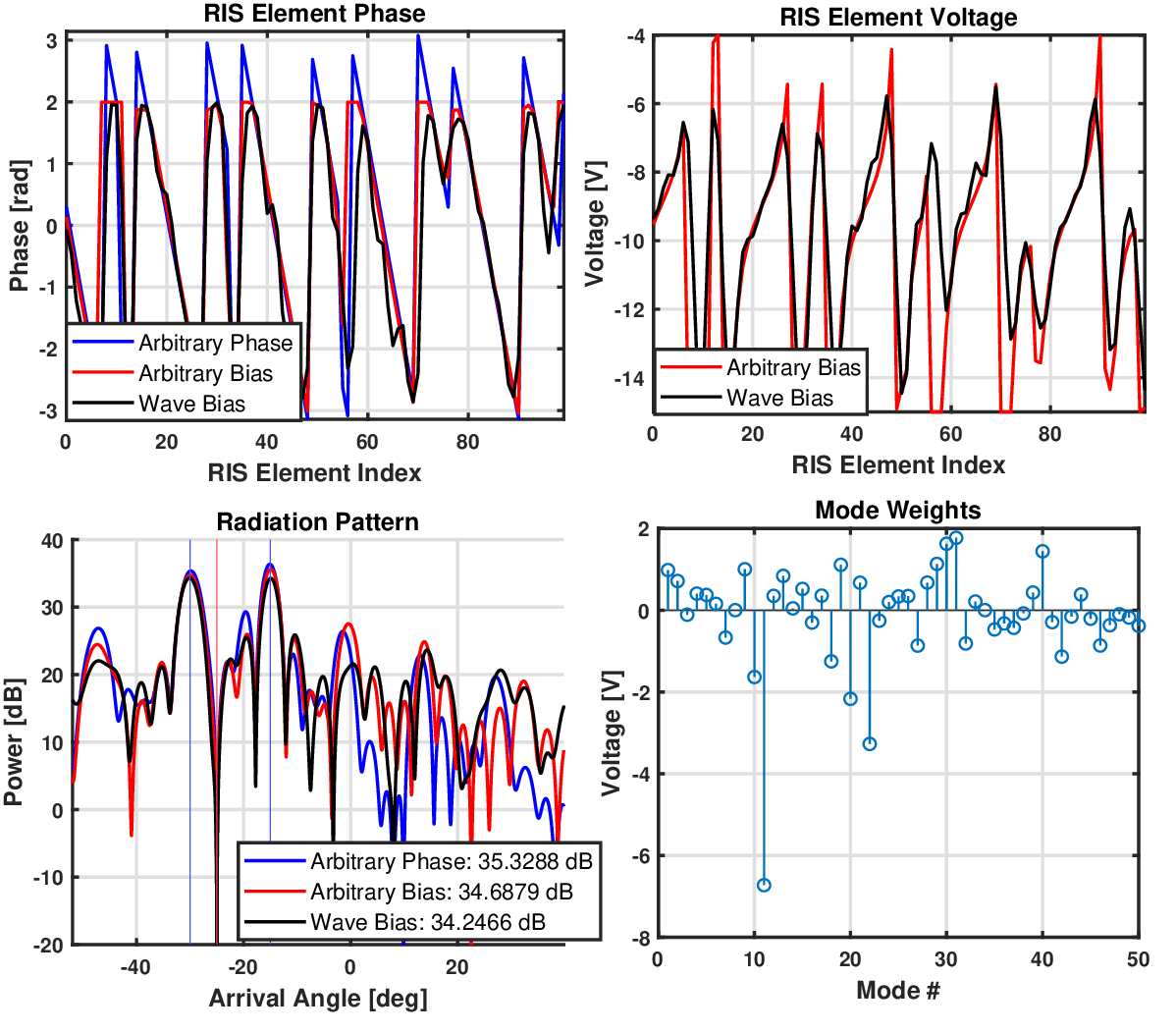}
    \caption{Simultaneous beam- and null-steering using the sample-and-hold model with $M=100$ RIS elements, $N=50$ modes, $M_l=M_r=2$. Desired beams at $-30\degree$ and $-15\degree$ and one null at $-25\degree$.}
    \label{fig:slnr_final_100}
\end{figure}

\begin{figure}[!t]
    \centering
    \includegraphics[width=1\linewidth]{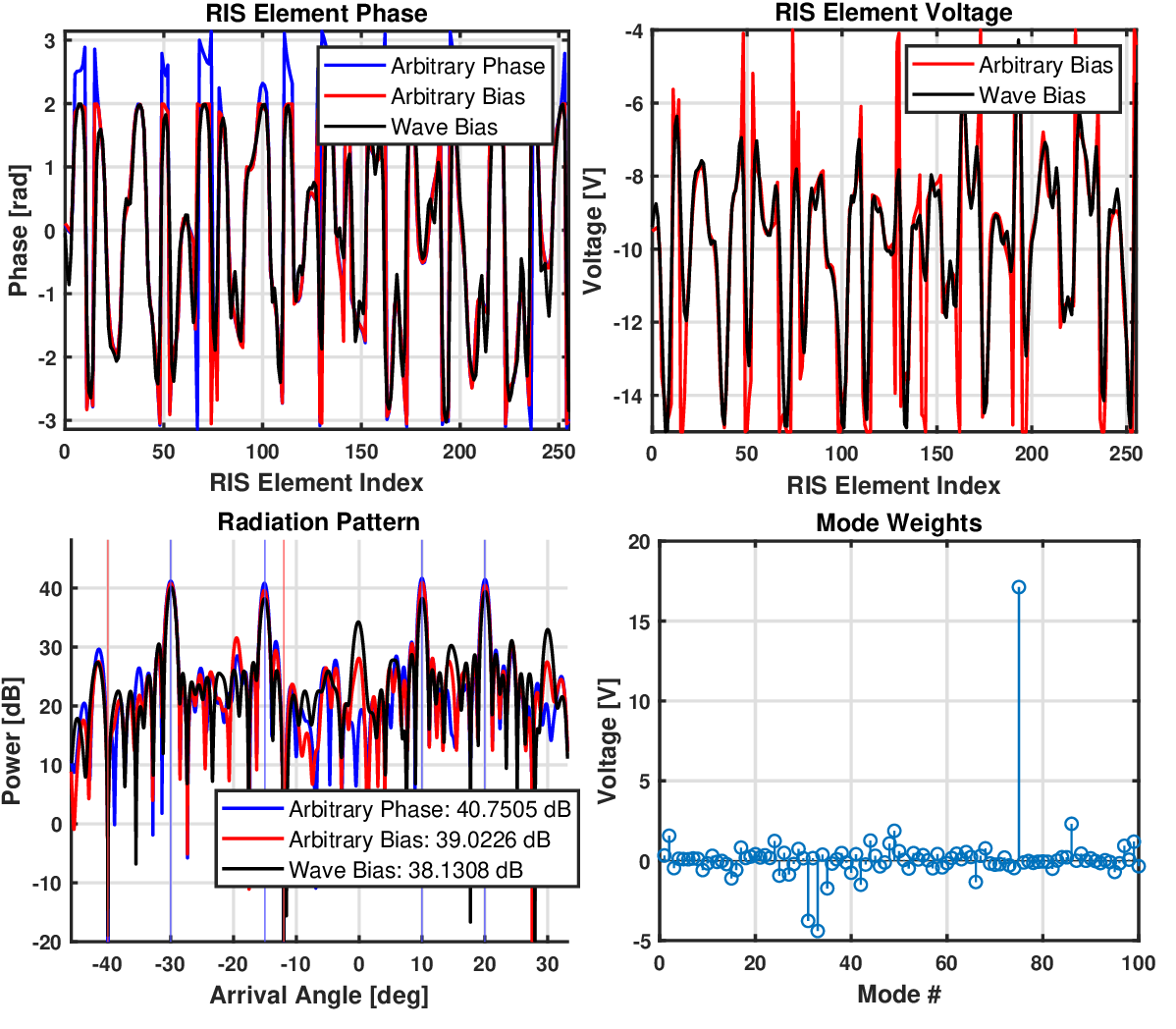}
    \caption{Simultaneous beam- and null-steering using the sample-and-hold model with $M=256$ RIS elements, $N=100$ modes, $M_l=M_r=2$. Desired beams at $-30\degree, -15\degree, 10\degree$, and $20\degree$, and nulls at $-40\degree$ and $-12\degree$.}
    \label{fig:slnr_final_256}
\end{figure}

Simulation results showing the performance of the above Combined Algorithm are given in Figs.~\ref{fig:slnr_final_100} and~\ref{fig:slnr_final_256}. We see that this algorithm implemented for the wave-controlled approach has less than 1 dB of loss in SLNR compared with the use of arbitrary biasing voltages for both cases. The beampatterns show strong peaks in the desired directions (blue vertical lines) and deep nulls in the undesired directions (red vertical lines). 

\subsubsection{SLNR Gain for Various Numbers of Elements and Modes}
\begin{figure}[!t]
    \centering
    \includegraphics[width=1.0\linewidth]{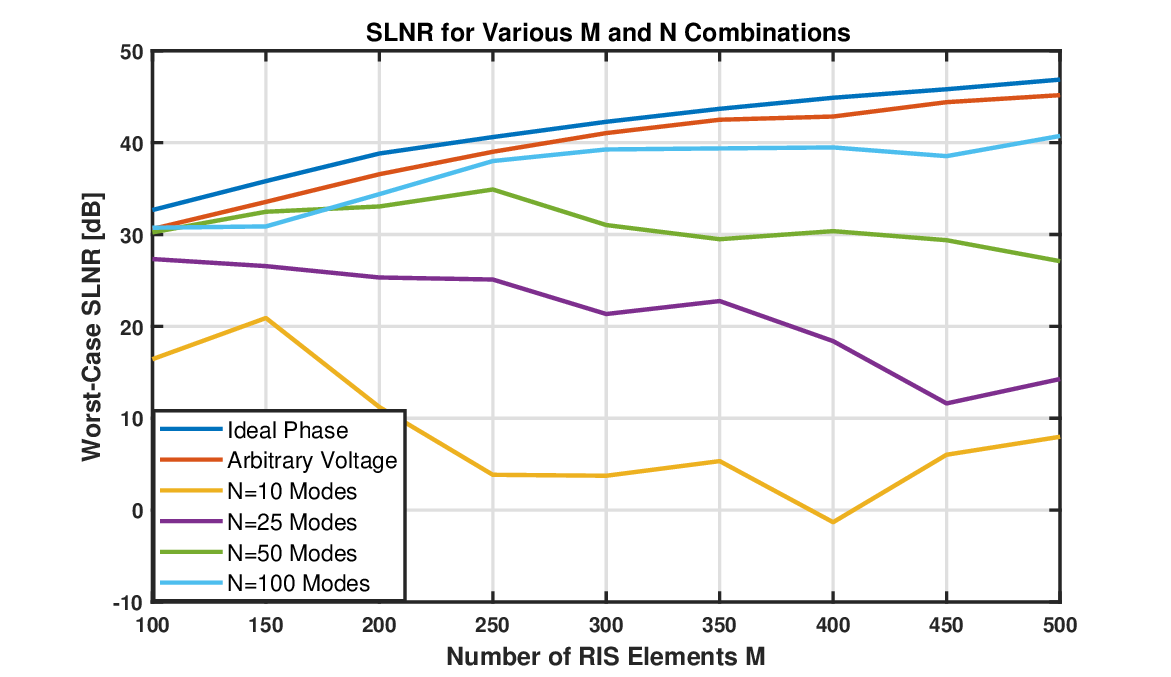}
    \caption{Worst-case SLNR for various numbers of RIS elements and modes with the sample-and-hold circuit. There are desired beams at $-30\degree, -15\degree, 10\degree$ and $20\degree,$ and nulls at $-12\degree$ and $-40\degree$. Each data point is the result of optimization using the Combined Algorithm, averaged over 10 trials.}
    \label{fig:slnr_vs_m_n}
\end{figure}

In this section, using the Combined Algorithm, we compare the performance of the proposed waveguide RIS for different numbers of RIS elements and standing-wave modes, using the sample-and-hold circuit realization.
For the first scenario, we study performance versus the number of modes $N$, where in this case we use the first $N$ modes in the decomposition. 
The case considered is the same as in Fig.~\ref{fig:slnr_final_256}, with four desired beams at $-15\degree, -30\degree, 10\degree,$ and $20\degree$, and two nulls at $-12\degree$ and $-40\degree$. All RIS configurations are assumed to have a transmission line extended by $2d_x$ both on the left and right ($M_l=M_r=2$). The performance of the different RIS designs is plotted in Fig.~\ref{fig:slnr_vs_m_n}.

\begin{figure}[!t]
    \centering
    \includegraphics[width=1.0\linewidth]{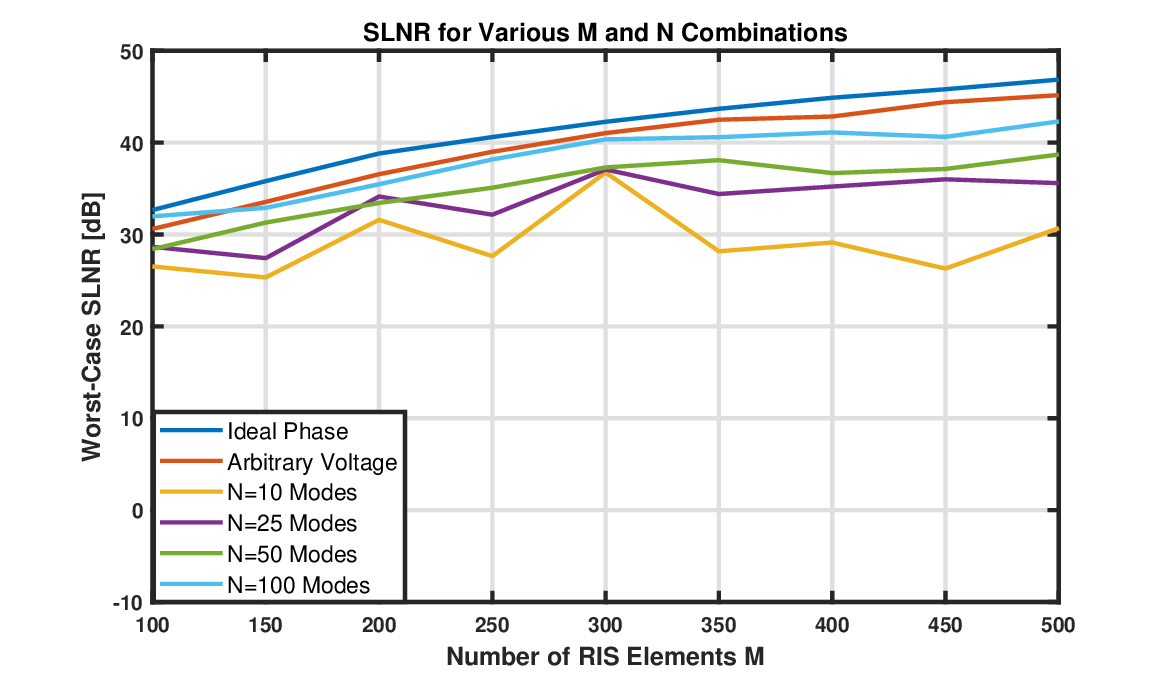}
    \caption{Worst-case SLNR for various numbers of RIS elements and modes with the sample-and-hold circuit, where only the modes with the strongest amplitudes are selected. There are desired beams at $-30\degree, -15\degree, 10\degree$ and $20\degree,$ and nulls at $-12\degree$ and $-40\degree$. Each data point is the result of optimization using the Combined Algorithm, averaged over 10 trials.}
    \label{fig:slnr_vs_m_n_2}
\end{figure}

It is observed that the SLNR performance of the ideal phase and arbitrary voltage cases grow steadily with the number of RIS elements. Interestingly, the SLNR for the wave-controlled approach only increases with $M$ when $N$ is large, due to the fact that we are using only the first $N$ harmonics, and small values of $N$ mean that the modes cover a relatively small and decreasing set of frequencies as $M$ grows. This is clear from the results in Fig.~\ref{fig:slnr_vs_m_n_2} for the same scenario, except in this case we choose the $N$ strongest modes to construct the wave-controlled biasing. Here we see that relatively few modes are needed to nearly match the performance achievable with arbitrary phase control, and that increases in the number of modes past a certain point provides a relatively marginal benefit.

\section{Conclusion}
This paper has presented implementation aspects associated with an RIS architecture in which the varactor biasing voltages are supplied by standing waves on a transmission line. The standing wave is created by a waveform generator that can control the amplitude of a number of harmonically related sinusoidal modes injected at one end of the transmission line. Such an architecture significantly simplifies the wiring and circuitry required to control the RIS, and potentially reduces the amount of control information that must be sent for RIS configuration. Two methods of converting the AC standing waves to DC varactor biasing voltages have been studied: an envelope detector and a sample-and-hold circuit. Based on models for these circuits, algorithms for optimizing the mode amplitudes have been developed to design radiation patterns with desired beam- and null-steering. While the envelope detector circuit is simpler to implement, optimization of the mode amplitudes is significantly more complicated and provides performance that is inferior to the sample-and-hold architecture. Simulations of the system performance demonstrate the ability of the wave-controlled RIS to generate strong beams and deep nulls in desired directions, with a relatively small degradation in terms of SNR or SLNR compared with the case of arbitrary control of each varactor element and idealized RIS models in which the RIS phase response can be arbitrarily specified.

\appendices
\section{Derivation of LS Results}
We wish to replicate an arbitrary voltage waveform $V(m)$ for all $m=0,1,...,M-1$ using $N$ sinusoids that construct the waveform $w(m)$ as in (\ref{eq:mmse_wave_model_dummy_elements}) and using DC voltage bias $W_0$ given by (\ref{eq:mmse_w0}).
The variable $w(m)$ can be re-written as
\begin{equation}
    w(m)=W_0+\boldsymbol{W}^T \boldsymbol{s}_m
\end{equation}
where $\boldsymbol{W}=[W_1, W_2, \ldots, W_N]^T$ is the vector representing all the mode weights, and
\[
\begin{split}
\boldsymbol{s}_m= \Big[
\sin & \left(\frac{\pi (m+M_l)}{M-1+M_l+M_r}\right)\sin(\omega_b t_0),\dots , \\
&
 \sin\left(\frac{N\pi (m+M_l)}{M-1+M_l+M_r}\right)\sin(N\omega_b t_0)
\Big]^T
\end{split}
\]
is the vector containing all the sinusoid terms before they are multiplied by each mode weight, at any RIS element $m$.
Define the cost function to minimize as
\begin{equation}
    J = \sum_{m=0}^{M-1}{||w(m)-V(m)||^2_2} .
\end{equation}
Expanding, we have
\begin{equation}
    \min_{\boldsymbol{W}}{J} = \min_{\boldsymbol{W}}\sum_{m=0}^{M-1}{\left(V^2(m)-2V(m)w(m)+w^2(m)\right)} .
\end{equation}
To minimize the cost function, take its gradient or vector derivative and set it equal to the zero vector
\begin{equation}
    \frac{\partial J}{\partial \boldsymbol{W}} = \sum_{m=0}^{M-1}{\left(-2V(m)\frac{\partial w(m)}{\partial \boldsymbol{W}}+2w(m)\frac{\partial w(m)}{\partial \boldsymbol{W}}\right)}= {\bf 0}
\end{equation}
where we use the notation $\frac{\partial J}{\partial \boldsymbol{W}}$ to mean the gradient of $J$ with respect to the vector $\boldsymbol{W}$.
The partial derivative of $w(m)$ with respect to $\boldsymbol{W}$ is
\begin{equation}
    \frac{\partial w(m)}{\partial \boldsymbol{W}}=\boldsymbol{s}_m .
\end{equation}
Plugging back, we get
\begin{equation}
    \frac{\partial J}{\partial \boldsymbol{W}} = \sum_{m=0}^{M-1}{\left(-2V(m)\boldsymbol{s}_m+2[W_0+\boldsymbol{W}^T \boldsymbol{s}_m]\boldsymbol{s}_m\right)}= {\bf 0} .
\end{equation}
Since $\boldsymbol{W}^T \boldsymbol{s}_m$ is a constant, it is equivalent to its transpose $\boldsymbol{s}_m^T\boldsymbol{W}$. Also, vectors can be multiplied by constants from either side, therefore
$
\boldsymbol{W}^T\boldsymbol{s}_m^{}\boldsymbol{s}_m^{} = \boldsymbol{s}_m^{}\boldsymbol{s}_m^T\boldsymbol{W},
$
and
\begin{equation}
    2\sum_{m=0}^{M-1}{\boldsymbol{s}_m^{}\boldsymbol{s}_m^T\boldsymbol{W}}=2\sum_{m=0}^{M-1}{\left(V(m)-W_0\right)\boldsymbol{s}_m} .
\end{equation}
Solving for optimal $\boldsymbol{W}$ yields
\begin{equation}
    \boldsymbol{W}=\left(\sum_{m=0}^{M-1}\boldsymbol{s}_m^{} \boldsymbol{s}_m^T\right)^{-1}
\left(\sum_{m=0}^{M-1}{(V(m)-W_0)\boldsymbol{s}_m}\right) .
\end{equation}
This expression will yield the minimum of the cost function due to the positive definite nature of the $\sum_{m=0}^{M-1}{\boldsymbol{s}_m^{}\boldsymbol{s}_m^T}$ matrix, which also allows its inversion \cite{P-probability}.\\

\noindent\textit{Theorem:}
The $\boldsymbol{s}_m$ vectors are linearly independent, immediately implying that the matrix
sum $\sum_{m=0}^{M-1}{\boldsymbol{s}_m^{}\boldsymbol{s}_m^T}$ is positive definite.\\
\textit{Proof:}
To prove that $\sum_{m=0}^{M-1}\boldsymbol{s}_m\boldsymbol{s}_m^T$ is positive definite, we will first prove, in items 1) and 2) below, that $\{\boldsymbol{s}_m\}_{m = 0}^{M-1}$
is a linearly independent set under the given conditions, and then, in item 3) below, we will prove that $\sum_{m=0}^{M-1}\boldsymbol{s}_m\boldsymbol{s}_m^T$ is
positive definite.
\begin{enumerate}
\item Each $\boldsymbol{s}_m^{}$ is generated using sinusoids of the form
$\sin\left(\frac{n\pi(m+M_l)}{M-1+M_l+M_r}\right)$, multiplied by weighting factors
$\sin(n\omega_b t_0)\neq0$ for every $n=1,2,\ldots,N$. Assume $M_l=0$ and
$M_r=0$. Then, the maximum number of $N$ for which
$\boldsymbol{s}_m^{}\neq\boldsymbol{0}$ is $M-2$ (due to the cases where $m=0$ and $m=M-
1$, because $\sin(0)=\sin(\pi)=0$). To get additional contributions from the edge cases for the
LS solution, it is sufficient to have $M_l \geq 1$ and $M_r \geq 1$.
\item Since each $\boldsymbol{s}_m^{}$ has sinusoidal components with frequencies
dependent on $m$, each $\boldsymbol{s}_i^{}$ is linearly independent from
$\boldsymbol{s}_j^{}$ where $i\neq j$. The proof is given below.\\
\textit{Lemma:} $N$ signals are linearly independent in the time domain if and only if they are
linearly independent in the frequency domain.\\
\textit{Proof:} The set of functions $\{g_i(t)\}_{i=1}^N$ is linearly independent on
$(-\infty, \infty)$ if
\begin{equation}
\sum_{i=1}^N a_i g_i(t) = 0,\quad t \in (-\infty, \infty)
\end{equation}
implies $a_i = 0$ for $i = 1,2,\ldots, N$ \cite{functional-analysis}. Assume there exist
constants $a_i$ for which $\sum_{i=1}^N a_i g_i (t) = 0$. Taking the Fourier transform results in
\begin{equation}
\begin{split}
\mathscr{F} \left\{ \sum_{i=1}^N a_i g_i(t) \right\} & = \sum_{i=1}^N a_i \mathscr{F}\{ g_i
(t) \} \\
& = \sum_{i=1}^N a_i G_i(f) = 0
\end{split}
\end{equation}
since $\mathscr{F} \{0\} = 0$. In other words, the same set of $a_i$ makes the linear
combination in the frequency domain equal to zero. If the functions $g_i(t)$ all have different
frequency components, they will all occupy separate sections in the frequency domain.
Therefore, the summation of them will only amount to zero for $f\in(-\infty,\infty)$ if
$a_i=0$ for $i=1,2,\ldots,N$. This argument shows that linear independence in the time domain
implies linear independence in the frequency domain (sufficient condition).
The necessary condition follows from the duality property of the Fourier transform.
\item
\textit{Lemma:} If $\{\boldsymbol{s}_m\}_{m=0}^{M -1}$ is a linearly independent set of vectors,
then the matrix $\sum_{m=0}^{M-1}{\boldsymbol{s}_m\boldsymbol{s}_m^T}$ is positive definite.\\
\textit{Proof:} Assume $\sum_{m=0}^{M-1}{\boldsymbol{s}_m\boldsymbol{s}_m^T}$ is not positive definite. Then, there exists a vector $\boldsymbol{w}_0$ such that $\boldsymbol{w}_0^T \left(\sum_{m=0}^{M-1}{\boldsymbol{s}_m\boldsymbol{s}_m^T}\right)\boldsymbol{w}_0$ is not greater than 0. Consider
\begin{eqnarray}
\boldsymbol{w}_0^T \left(\sum_{m=0}^{M-1}\boldsymbol{s}_m\boldsymbol{s}_m^T\right)\boldsymbol{w}_0 & = & \sum_{m=0}^{M-1} \boldsymbol{w}_0^T\boldsymbol{s}_m\boldsymbol{s}_m^T\boldsymbol{w}_0, \nonumber\\
& = & \sum_{m=0}^{M-1}(\boldsymbol{w}_0^T \boldsymbol{s}_m)^2 . \label{eqn:sumofsquares}
\end{eqnarray}
The quantity in (\ref{eqn:sumofsquares}) is a sum of squares, therefore it cannot be less than zero. So, if $\sum_{m=0}^{M-1}\boldsymbol{s}_m\boldsymbol{s}_m^T$ is not positive definite, then $\sum_{m=0}^{M-1}(\boldsymbol{w}_0^T\boldsymbol{s}_m)^2=0.$ This can only happen if $\boldsymbol{w}_0^T\boldsymbol{s}_m = 0$ for all $m=0,1,\ldots,M-1$. But, that means $\boldsymbol{s}_m$ are all proportional, i.e.,
\[
\boldsymbol{s}_m = \beta_m \boldsymbol{s}_0\quad m=0,1,\ldots, M-1,
\]
where $\beta_m$ is a constant.
Which implies $\{\boldsymbol{s}_m\}_{m=0}^{M-1}$ is not a linearly independent set. But that contradicts the hypothesis and the proof is complete.
\end{enumerate}

The summation of $M$ such $(N\times N)$ matrices therefore results in a full rank matrix with
nonzero eigenvalues, where $N\leq M-2+\min{(M_l, 1)} + \min{(M_r, 1)}$. Therefore,
$\sum_{m=0}^{M-1}{\boldsymbol{s}_m^{} \boldsymbol{s}_m^T}$ is positive definite and
invertible. The matrix would remain positive definite also for the case $\sum_{m=0}^{M-
1}{\alpha(m) \boldsymbol{s}_m^{} \boldsymbol{s}_m^T}$ where $\alpha(m) > 0$, since these are
just scaling factors that would not interfere with the number of positive eigenvalues in the overall
summation matrix.

\section{Main Modes Corresponding to Specific Reflection Angles}
Let $\boldsymbol{W}=\left[0, 0, \ldots, 0\right]^T$ be the $N\times 1$ zero vector containing all vanishing mode amplitudes,  except for one index $n$. The resulting standing wave voltage at each element $m=0,\ldots,M-1$, is given by
\begin{equation}
    \begin{split}
        w(m) & = W_0+W_n \sin\left(\frac{n\pi m}{M-1}\right)\sin(n\omega_b t_0) \\& =
        W_0 + C \sin\left(\frac{n\pi m}{M-1}\right).
    \end{split}
\end{equation}
This means that $w(m)$ oscillates with a spatial angular frequency $\kappa=\frac{n\pi}{M-1}$. This also suggests that the phase shift $\varphi\left(w(m)\right)$ of the RIS reflection coefficient  will also oscillate with that same spatial frequency since the conversion between voltage to phase is one-to-one (although it is nonlinear), for the frequencies of interests shown in Fig.~\ref{fig:Phase_Ref_Coef}. While still maintaining the assumption that the amplitudes of the reflection coefficients are $|\phi(m)|\leq 1$, one can approximate
\begin{equation}
    \phase{\phi(m)}\approx \phi_0 + D\sin\left(\frac{n\pi m}{M-1}+\alpha\right).
    \label{eq:appxB_phase}
\end{equation}
Assume $\phi_0 = 0$, $\alpha=0$, and $D=1$ for simplicity.
The power directed towards a specific receiver direction $\theta^*$ is calculated using the definitions from Section V-A,
\begin{equation}
    P=\rho_s \left|\sum_{m=0}^{M-1} {\phi(m) e^{-jm\kappa(\theta^*)}}\right|^2.
\end{equation}
Applying Euler's identity to~(\ref{eq:appxB_phase}) and combining with the above definition gives
\begin{equation}
    \begin{split}
        P & \approx \rho_s \left| \sum_{m=0}^{M-1} {\frac{1}{j2}\left[e^{\left(\frac{jn\pi m}{M-1}\right)} - e^{\left(-\frac{jn\pi m}{M-1}\right)}\right]e^{-jm\kappa(\theta^*)}}\right|^2 \\
        & =
        \rho_s \left| \frac{1}{2} \sum_{m=0}^{M-1}e^{\left[ j\left(\frac{n\pi m}{M-1}-m\kappa(\theta^*)\right)\right] } - e^{\left[-j\left(\frac{n\pi m}{M-1}+m\kappa(\theta^*) \right) \right]} \right|^2.
    \end{split}
\end{equation}
Since $n>0$, only the first complex exponential term can become unity to maximize the power towards $\theta^*$, thus
\begin{equation}
    \frac{n\pi m}{M-1}=m\kappa(\theta^*).
\end{equation}
Without the case where $m=0$, the index $n$ that maximizes the power is
\begin{equation}
    n=\frac{(M-1)\kappa(\theta^*)}{\pi}.
    \label{eq:appendix_2_n}
\end{equation}
Substituting $\kappa(\theta^*)$ as $2\pi\Delta\sin(\theta^*)$ in~(\ref{eq:appendix_2_n}) and taking the absolute value since $n>0$ results in
\begin{equation}
    n=\left| \frac{2\pi\Delta(M-1)\sin(\theta^*)}{\pi} \right| = \left| 2(M-1)\Delta\sin(\theta^*) \right|,
\end{equation}
which can be rounded to the nearest integer value $\round{\cdot}$. 
\bibliographystyle{IEEEtran}
\bibliography{IEEEabrv,bibJournalList,bibfile,refs}
\vfill
\end{document}